\documentclass[a4paper,fleqn,usenatbib,useAMS]{mnras}

\usepackage{newtxtext}

\usepackage[T1]{fontenc}
\usepackage{ae,aecompl}

\usepackage[pdftex]{graphicx}
\usepackage{amsmath}	
\usepackage{indentfirst}	
\usepackage{pict2e}
\usepackage{color}
\usepackage{bm}
\usepackage{tablefootnote}
\usepackage{pict2e}

\usepackage{ulem}
\newcommand{\adr}[1]{\textcolor{black}{ #1}}
\newcommand{\adm}[1]{\textcolor{black}{ #1}} 
\newcommand{\adc}[1]{\textcolor{black}{ #1}} %
\newcommand{\adrr}[1]{\textcolor{black}{ #1}}

\usepackage{wrapfig}
\graphicspath{{./figs/}}
\usepackage{color}
\usepackage{amsmath}
\usepackage{amssymb}

\newcommand{\Msun}{{\rm  M_{\odot}}}

\newcommand{\Zsun}{Z_{\odot}}

\newcommand{\solmass}{\rm M_{\odot}}

\newcommand{\Msunyr}{M_\odot~{\rm yr}^{-1}}

\providecommand{\adsurl}[1]{\href{#1}{ADS}}

\newcommand{\Hm}{\rm H^-}
\newcommand{\HH}{\rm H_2}

\newcommand{\diff}[3]{\mathinner{\frac{d ^{#1} {#2}}{d {#3} ^{#1}}}}

\title[Formation of first galaxies] 
{\adc{Formation of the first galaxies in the aftermath of the first supernovae}
}
\author[Abe et al. ]{Makito Abe$^{1}$\thanks{E-mail: mabe@ccs.tsukuba.ac.jp}, Hidenobu Yajima$^{1}$, Sadegh Khochfar$^{2}$, Claudio Dalla Vecchia$^{3}$ \newauthor and Kazuyuki Omukai$^{4}$ \\
$^{1}$Center for Computational Sciences, University of Tsukuba, Ten-nodai, 1-1-1 Tsukuba, Ibaraki 305-8577, Japan\\
$^{2}$Institute for Astronomy, University of Edinburgh, Royal Observatory, Edinburgh, EH9 3HJ, UK\\
$^{3}$Instituto de Astrof\'isica de Canarias, C/V\'ia L\'actea s/n, 38205 La Laguna, Tenerife, Spain\\
$^{4}$Astronomical Institute, Graduate School of Science, Tohoku University, Aoba, Sendai 980-8578, Japan}

\begin{document}

\date{Accepted ?; Received ??; in original form ???}

\pagerange{\pageref{firstpage}--\pageref{lastpage}} \pubyear{2008}

\maketitle

\label{firstpage}

%
%
\begin{abstract}
We perform high-resolution cosmological hydrodynamic simulations to study the formation of the first galaxies that reach the masses of $10^{8-9}~h^{-1}~M_\odot$ at $z=9$. 
The resolution of the simulations is high enough to resolve minihaloes 
and allow us to successfully pursue the formation of multiple \adc{Population (Pop)} III stars, their supernova (SN) explosions, resultant metal-enrichment of the inter-galactic medium (IGM) in the course of the build-up of the system. 
Metals are ejected into the IGM by multiple Pop III SNe, but
some of the metal-enriched gas falls back onto the halo after $\gtrsim 100~\rm Myr$. 
\adr{The star} formation history \adr{of} 
the first galaxy depends sensitively on the initial mass function (IMF) of Pop III stars. 
The dominant stellar population transits from Pop III to Pop II at $z\sim 12-15$ in the case of power-law Pop III IMF, ${\rm d}n/{\rm d}M \propto M^{-2.35}$ with the mass range $10-500~\Msun$. 
\adc{At $z\lesssim 12$, stars are stably formed in the first galaxies with a star formation rate of $\sim 10^{-3}$-$10^{-1}~M_\odot/{\rm yr}$. }
\adc{In contrast,} for the case with a flat IMF, 
\adc{gas-deprived} first galaxies form due to frequent Pop III pair-instability SNe, resulting in the suppression of subsequent Pop II star formation. 
In addition, we calculate UV continuum, Ly$\alpha$- and H$\alpha$-line fluxes from the first galaxies.  We show that the {\it James Webb Space Telescope} will be able to detect both UV continuum, Ly$\alpha$ and H$\alpha$ line emission from first galaxies with halo mass $\gtrsim 10^{9}~\Msun$ at $z \gtrsim 10$. 

\end{abstract}
%
%

%
%
\section{Introduction}
In the modern cosmology, first stars, so-called Population III (Pop III) stars, are supposed to form in small H$_2$-cooling haloes, so-called minihaloes ($\sim 10^6 M_\odot$) at $z\sim 10-30$ \citep[e.g.,][]{Couchman1986, Haiman1996,Tegmark+97, Abel2002, Bromm2002, Yoshida+03}. 
Star formation in minihaloes occurs presumably just in one or two episodes because of fragility of molecular hydrogen and strong radiative/supernova feedback exerted by the formed Pop III stars \citep{Omukai1999, Glover2001}. 
Through \adr{the} hierarchical merging process, those small haloes subsequently grow into a larger system. 
Once their mass exceeds $\sim 10^{7}~\Msun$ and the virial temperature reaches $\sim 10^{4}~\rm K$, 
\adc{the gas in} the halo collapses via more robust hydrogen atomic cooling, 
\adc{and the hosting halo is referred to as } an atomic cooling halo 
\adc{and the galaxy as a } {\it first galaxy}, where star formation proceeds in a sustainable way \citep[e.g.,][]{Bromm&Yoshida11}. 

The first galaxies are the initial stage of galactic evolution and building blocks of the present-day counterparts. 
Also, they are believed to be important radiation sources in the cosmic reionization \citep{Yajima+11,Yajima+14,Wise+12,Paardekooper+13,Kimm&Cen14,Paardekooper+15,Trebitsch+18,Ma+20arXiv}.
Understanding the formation of the first galaxies is one of major goals in contemporary 
\adc{extra-galactic} astronomy. 
Observationally, the first galaxies are among prime targets in the era of next generation telescopes, such as 
the  {\it{James~Webb~Space~Telescope}} ({\it{JWST}}). 
Thus, theoretical prediction for their physical properties is vital for successful 
\adc{searches} with future observations.

Pop III stars are thought to form by gravitational collapse induced by $\HH$ cooling \citep[e.g.,][]{Omukai1998, Yoshida2008}.
Theoretical studies 
\adc{assessing the initial mass function (IMF) of Pop III stars predict their masses widely ranging in $\sim$10-1000~$\solmass$}
\citep[e.g.,][]{Nakamura&Umemura01, Hirano+14, Susa+14, Hirano+15}. 
At the end of their lives, Pop III stars with the initial mass of 8-40~$M_\odot$ are expected to explode as core-collapse supernovae (CCSNe). If stars are in the mass range of 140-260~$M_\odot$, they explode as pair-instability SNe (PISNe). 
\adc{More massive stars} directly collapse into black holes (BHs) without explosion \citep{Heger&Woosley02}. 
The supernovae evacuate the gas not only in the host but also in neighbouring haloes, thereby suppressing subsequent star formation in them 
{\citep[e.g.,][]{BrittonSmith+15,Chiaki+18}}. 
Furthermore, massive Pop III stars emit intense ultraviolet (UV) radiation \citep{Tumlinson2000, Bromm01, Schaerer02}. 
The ionizing and H$_2$-dissociating (so-called Lyman-Werner, LW) photons prevent the collapse of gas and subsequent star formation not only in the parental minihaloes \citep{Omukai1999,Glover2001} but also in neighbouring ones \citep{Haiman+1997, Susa&Umemura06, Hasegawa+09,Trenti&Stiavelli09,Agarwal+14,Latif&Khochfar19}. 
This radiative feedback 
\adc{has been shown to play a major role} in regulating the cosmic Pop III star formation rate density 
\adc{in the Universe} \citep{FiBY}. 

As for modelling of first galaxy formation, \citet{Greilf+10} carried out high-resolution cosmological hydrodynamics simulations considering the Pop III SNe. 
They showed that although the Pop III SN explosion ejects metal-enriched gas into the surrounding IGM, the ejected gas eventually falls back onto 
the halo as it evolves into a larger system, with its mass $M_{\rm vir}=10^8~\solmass$ at $z=10$. 
If a single PISN occurs, metallicity at the centre of the halo can reach $Z\sim 10^{-3}~Z_\odot$, which is sufficient to cause the transition of star formation mode from massive Pop III to low-mass Pop II stars \citep{Omukai2000, Bromm01, Schneider2002, Omukai+05}.
\citet{Pawlik+13} carried out radiation-hydrodynamic simulations to investigate the impact of photo-ionization/dissociation radiation during  the formation of the first galaxies. 
They pointed out that  radiation feedback can suppress condensation of the gas and star formation in the main progenitor minihalo.
\citet{Jeon+15}  investigated the assembly of a first galaxy that reaches the mass of $\sim 10^8~M_\odot$ at $z\sim 10$ assuming a top-heavy initial mass function (IMF) for Pop III stars. 
\citet{FiBY} performed a high-resolution large-scale cosmological hydrodynamic simulation taking into account the photo-dissociating radiation and supernovae (SNe) feedback in modelling Pop III star formation, first metal enrichment and formation of Pop II stars in galaxies.
Taking advantage of the large ($>$~Mpc on a side) simulation box, they could pin down the star formation rate (SFR) densities of Pop II and III stars, and the transition redshift from Pop III to Pop II stars. 
However, 
\adc{those studies have not resolved formation of individual Pop III stars but only considered average IMF properties}. 

On the other hand, high-resolution hydrodynamic simulations 
with smaller computational boxes resolving a single Pop III star were carried out to see its impact on formation of second-generation stars. 
\citet{Chiaki+18} investigated metal enrichment due to Pop III SNe by way of cosmological radiation hydrodynamics combined with a metal propagation model based on a Lagrangian test-particle technique. 
They showed that SN ejecta fall back to the original halo and the internal enrichment takes place in the case of a CCSN, whereas the ejecta reach a neighbouring minihalo and induces its external enrichment in the case of PISN explosion \citep[see also,][]{Maio+11,BrittonSmith+15,Hicks+20}.  
Due to \adr{the} numerical cost in resolving the detailed structures of minihaloes, however, their simulation was limited to a small volume containing only several minihaloes. 

Thus, building-up of a first galaxy,  starting from the formation of individual Pop III stars,  has not been 
\adc{modelled in great detail yet.} 
One exception is the {\it Renaissance} simulations, where high-resolution adaptive mesh refinement (AMR) cosmological hydrodynamic simulations have been pioneered with modelling Pop III stars and first galaxies \citep{Xu+13,Xu+14,Chen+14,Ahn+15,O'Shea+15}.  
They focused on overdense regions and investigated the evolution up to $z = 15$. 
More than $10,000$ Pop III stars and $\sim 1,000$ first galaxies are found in their computational volume. 
Although they also followed formation of individual Pop III stars, the minimum halo mass they resolved is limited to $\sim 10^6~M_\odot$ and stars formed in smaller haloes were not properly taken into account. 
\adr{In addition, in the modelling of the Pop III star formation, they assumed a single functional form for the IMF although it is still under debate \citep[e.g,][]{Nakamura&Umemura01,Susa+14,Hirano+15,Stacy+16}. 
It is still unclear how 
the IMF of the Pop III stars impacts on the physical properties of the first galaxies. }
In this paper, we perform high-resolution cosmological hydrodynamic simulation, resolving individual Pop III stars in minihaloes \adr{down to $\sim 10^{5}~M_\odot$} and its subsequent evolution, and investigate the impact of multiple Pop III feedback events on the formation of the first galaxies. 
\adr{Besides, we investigate the effects of different Pop III IMFs by changing the power-law slope of the IMF. }

This paper is organized as follows. 
In Section 2, we present the numerical method for our simulations. 
The numerical results are described in Section 3, where we show the assembly of first galaxy taking into account the mass dependence of the system. 
Effects of different model assumptions on e.g., treatment of the LW radiation feedback, adopted Pop III IMF, etc.  
are discussed in Section 4. 
Section 5 is devoted to discussion on observability of first galaxies by future instruments. 
Finally, we conclude our results in Section 6. 
Throughout this paper, we assume $\Lambda$CDM cosmology with the cosmological parameters; 
$\Omega_{\rm M} = 0.3$, $\Omega_{\rm b} = 0.045$, $\Omega_{\rm \Lambda} = 0.7$ and $h = 0.7$. 

%
%
\section{Method}
\subsection{Code and initial condition}
We carry out a suite of first galaxy simulations with the modified version of the smoothed particle hydrodynamic (SPH)/ $N$-body code GADGET-3 \citep{GADGET01, GADGET2}. 
The code treats the formation/evolution of Pop III stars, LW radiation feedback, SN feedback, non-equilibrium primordial chemistry with equilibrium metal cooling, which are implemented in the First Billion Year (FiBY) project \citep{FiBY}. 
Recently the code was updated in regard to the feedback from stars and black holes and the dust destruction in Forever22 project \citep{F22}.
Our implementation for the thermal evolution is based on \citet{FiBY}. 
We solve non-equilibrium primordial chemistry and molecular cooling \adr{and (collisional or photoionization) equilibrium metal cooling. }

In this work, we newly incorporate the photo-ionization feedback into the code. 
For the high-resolution cosmological hydrodynamic simulation, we generate a zoom-in initial condition by utilizing MUSIC code \citep{MUSIC}. 
First we carry out a cosmological $N$-body simulation with the box size of (4~$h^{-1}$~Mpc)$^3$ with $128^3$ coarse-grained dark matter particles. 
Then, we perform friends-of-friends (FOF) group finding to identify dark matter haloes with masses of $M_h \sim 10^{8}~h^{-1}~M_\odot$ (M8run) and  $M_h \sim 10^{9}~h^{-1}~M_\odot$ (M9run). 
We refer to M8run as our {\it fiducial run} and investigate its feedback and parameter dependencies. 
In addition, we carry out five more simulations with different initial conditions producing $\sim 10^{8}~\Msun$ haloes at $z=9$ considering diversity of galaxy formation environments (summarized in Table \ref{table:parameter}).  
The zoom-in regions have the size of $\sim 200-300$~comoving kpc, 
and contain high-resolution DM and SPH particles with the mass of  $m_{\rm DM}\sim 66~h^{-1}~\solmass$ and $m_{\rm SPH} \sim 12~h^{-1}~\solmass$, corresponding to the effective resolution of (4096)$^3$. 
We set the gravitational softening length $\adrr{\epsilon_{\rm g}} = 90$ comoving pc for both DM and SPH particles. 
\adm{We also consider a case of formation of a first galaxy in a higher-density region (M9runH). 
In the M9runH, we select a region that evolves into a $\sim 3\times 10^{9}~h^{-1}~M_\odot$ halo at $z = 9$ within a (14~$h^{-1}$~Mpc)$^3$ box and generate the zoom-in initial condition. 
}
In the \adm{M9runH}, the mass resolution is worse compared to M8run and M9run due to the computational cost. 
The refined DM and SPH particle masses are $m_{\rm DM}\sim 330~ h^{-1}~\solmass$ and $m_{\rm SPH} \sim 60 ~ h^{-1}~\solmass$, respectively.  
However, the mass resolution is still high enough to allow us to resolve minihaloes hosting Pop III stars.

\subsection{Star formation}
With resolution of physical $\sim$ pc scale, our simulations can follow the prestellar collapse of gas in minihaloes.
However, even with current computer facilities, it is too expensive to resolve individual protostar formation simultaneously with cosmological large-scale structure.
We avoid this difficulty by introducing a sink particle method, in which we convert gas particles into a collisionless single star particle, if the hydrogen number density $n_{\rm H}$ exceeds the threshold density of $10^3~{\rm cm^{-3}}$ and the velocity is converging $\nabla \cdot \vec{v}<0$. 
Recent cosmological simulations show that Pop III stars are typically more massive than $\sim 10~\Msun$ with the mass spectrum ranging from $\sim 10$ to $\sim 500~\Msun$ \citep[e.g.,][]{Hirano+15}. Note, however, that the \adrr{mass spectrum and its lower mass limit are} still under debate because of the difficulties in following long-term disk instability and fragmentation \citep[e.g.,][]{Clark+11,Clark+11Sci,Susa+14,Sugimura+20}.
For the fiducial cases shown in Section 3, we adopt a Salpeter-like IMF, e.g., 
$dn/dM \propto M^{-\alpha}$ with $\alpha
= 2.35$ with the mass range of $10 - 500~\Msun$, but also studied a case of more top-heavy IMF later in Section 4.
The mass of a single Pop III star particle is randomly chosen with probability given by the assumed IMF.
We note that if the stellar mass exceeds the SPH particle mass, we subtract the mass of the neighbouring gas particles to fill the gap. 

Pop II star formation takes place in the form of star clusters \citep[e.g.,][]{Lada&Lada03}. 
To take this into account, we treat each Pop II stellar particle as a ``star cluster'' and assume a single stellar population with the Chabrier IMF with mass in the range 0.1 - 100 $\Msun$. 
If $n_{\rm H} > 100~{\rm cm^{-3}}$ and $\nabla \cdot \vec{v}<0$ are satisfied, we estimate the local free-fall time $t_{\rm ff} = \sqrt{3\pi/32G\rho_{\rm gas}}$ which roughly corresponds to the local star-formation timescale. 
Then, the local star formation rate (SFR) is calculated as
\begin{equation}
	\diff{}{\rho_\ast}{t} = c_\ast \frac{\rho_{\rm gas}}{t_{\rm ff}} , 
	\label{eq:rho_star_dot}
\end{equation}
where $\rho_\ast$ is the local stellar density and $c_\ast$ is a dimensionless parameter that controls the star formation efficiency. 
Throughout this paper, we assume $c_\ast = 0.1$. 

\adr{
The thermal history of the collapsing gas sensitively depends on the metallicity, resulting in different stellar masses \citep{Omukai+05}. 
We assume that Pop III stars form if $Z < Z_{\rm cr}$, while Pop II star clusters form at higher metallicity. 
The critical metallicity is set to $Z_{\rm cr} = 10^{-4}~Z_\odot$ \citep[e.g.,][]{FiBY}. 
We also investigate the impacts of the critical metallicity in \S \ref{sec:critical_metallicity}.}

\subsection{Stellar feedback}
\subsubsection{Supernova and metal enrichment}
SN feedback alters structure of the inter-stellar/galactic media (ISM/IGM) significantly and thereby affects subsequent star formation. 
\adr{The} treatment of SN feedback is thus an essential element in studying the assembly of a first galaxy. 
In this study, we adopt the subgrid recipe proposed by \citet{DallaVecchia&Schaye12}, which can avoid the so-called overcooling problem regardless of the mass resolution. 
In this scheme, SN thermal energy is stochastically distributed to neighbouring SPH particles 
\adm{by elevating the temperature by $\Delta T$. }
We set the heating temperature $\Delta T = 10^{7.5}~$K and the SN energy $E_{\rm SN} = 10^{51} {\rm erg}$ per a supernova for CCSN of Pop III or Pop II, while $\Delta T = 3\times 10^{8.5}~$K and $E_{\rm SN} = 10^{53} {\rm erg}$ are set for PISN of Pop III. 
\citet{DallaVecchia&Schaye12} argued that thermal energy is efficiently converted into kinetic energy when the sound crossing time is less than the cooling time. 
This condition is satisfied for the gas density below the following value, 
\begin{equation}
	n_{\rm H} \sim 3000~{\rm cm^{-3}}~\left(\frac{\Delta T}{10^{7.5}K}\right)^{3/2}\left(\frac{m_{\rm SPH}}{10~\solmass}\right)^{-1/2}. 
	\label{eq:crit_dens}
\end{equation}
When SN releases the energy $E_{\rm SN}$, 
\adm{we randomly input the thermal energy to neighbouring SPH particles in accordance with the probability so that the total thermal energy coincides with $E_{\rm SN}$. } 
The probability $p_{\rm SN}$ is given by 
\begin{equation}
	p_{\rm SN} = \frac{f_{\rm th} E_{\rm SN}}{\Delta \epsilon \sum_i^{N_{\rm ngb}}m_i} 
	\label{eq:rho_star_dot}
\end{equation}
where $f_{\rm th}$ is the feedback efficiency and we assume it to be unity, $N_{\rm ngb}$ is the number of neighbour SPH particles and $m_i$ is the mass of an SPH particle 
\adrr{, $\Delta\epsilon = k_{\rm B}\Delta T/(\gamma-1)\mu m_{\rm H}$ is the increase of the thermal energy per unit mass. }
\adm{For each neighbour SPH particle, we generate the uniform random number $\xi$ between 0 and 1. 
We increase the temperature to $\Delta T$ if the SPH particle satisfies the condition of $\xi < p_{\rm SN}$. }
The expectation value of the number of heated particles is evaluated as 
\begin{equation}
	\langle N_{\rm heat} \rangle = \frac{f_{\rm th} E_{\rm SN}}{\Delta \epsilon}\frac{N_{\rm ngb}}{\sum_i^{N_{\rm ngb}}m_i} \sim  
						\frac{f_{\rm th} E_{\rm SN}}{m_i \Delta \epsilon}. 
	\label{eq:rho_star_dot}
\end{equation}
In this setup, we typically heat $\sim$ 5 particles for CCSN and $\sim$ 14 particles for PISN. 

\adrr{
We follow the metal enrichment as in \citet{Wiersma+09}. 
Once the the neighbour SPH particles are chosen for the thermal feedback, the metals produced by SNe are also injected into them. The mass of metals assigned to an SPH particle is estimated with the weight based on the kernel function. 
Note that, here we do not take into account the metal transfer between SPH particles. 
Instead, 
we consider the local metallicity smoothed over 48 neighbour SPH particles. 
}

\subsubsection{Photoionization feedback}

Previous studies pointed out that the ionization/dissociation radiation from Pop III stars regulates subsequent star formation nearby \citep[e.g.,][see also \citet{Susa+09, Hasegawa+09}]{FiBY, Pawlik+13}. 
However, the ionization radiation feedback has a strong effect only in early phases of the assembly and becomes powerless once the system reaches the atomic cooling regime \citep{Pawlik+13}. 
In systems lager than the atomic cooling halo limit, SN feedback becomes
\adc{more} important in regulating the evolution than radiation feedback \citep[e.g.,][]{Yajima+17}. 
In this work, we model the photo-ionized region based on a photon conservation approach as
\begin{equation}
	\dot{N}_{\rm ion} = \sum_i \frac{m_i}{\rho_i}\alpha_{\rm B} n_{{\rm H}, i}^2
	\label{eq:photoheating}
\end{equation}
where $m_i$, $\rho_i$, $n_{{\rm H}, i}$ are SPH particle mass, density and hydrogen number density of $i$-th particle, $\alpha_{\rm B}=2.59\times10^{-13}~{\rm cm^3~s^{-1}}$ is the case-B recombination coefficient at temperature $T=10^4~$K. 
Note that the summation is carried out in ascending order of distance from ionizing sources. 
For $\dot{N}_{\rm ion}$ of Pop III stars, we refer to the table in \citet{Schaerer02}. 
If SPH particles are in the photo-ionization region, the temperature is raised to $3 \times 10^{4}~\rm K$. 
The photo-ionization regions push the surrounding neutral gas via high  pressure even before the explosions of SNe. 
\adrr{In addition to Pop III stars, we also consider the photo-ionization feedback from Pop II stars in the same way. 
We set the ionizing photon emissivity per stellar mass for Pop II stars as $4\times 10^{46}~{\rm s^{-1}~M_{\odot}^{-1}}$, which is estimated from the SED of a typical high-redshift galaxy using {\sc starburst99} \citep{STARBURST99}. }

\subsubsection{$\HH$ photodissociation and $\Hm$ photodetachment}

We take into account UV radiation from young stars
to include the photodissociation of $\HH$ and photodetachment of $\Hm$. 
Since  $\HH$ molecules are the dominant coolant in low-metallicity gas at $T<10^4$~K \citep{Susa&Umemura2004}, 
UV radiation is likely to have a significant impact on Pop III/II formation efficiency and subsequent properties of the first galaxies. 
In calculating the dissociation rate of $\HH$, we evaluate the UV flux from the geometrical dilution $\propto L/r^{2}$, where $r$ is the distance from the radiation source to the gas, considering also the self-shielding based on the column density over a local Jeans length.
The self-shielding function proposed by \adrr{\citet[][]{Wolcott-Green+11} (equation 11 in their paper)} is used in our simulations. 

%
%

\section{Result}

\subsection{Formation of first galaxy} \label{sec:formation-evolution}

\begin{figure*}
	\begin{center}
		\includegraphics[width=14.0 cm,clip]{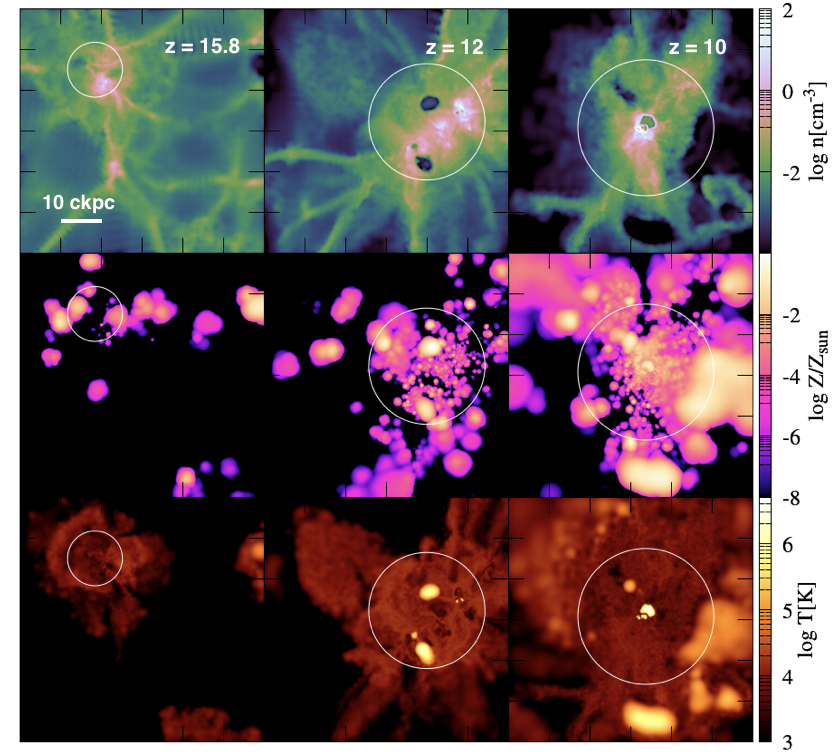}
	\end{center}
	\caption{
	Time evolution of density (top panel), metallicity (middle) and temperature distribution (bottom panel) on the x-y mid-plane in the case for {\it{fiducial run}}.  
	From left to right, the distributions are corresponding to redshift $z=15.8$, 12 and 10. 
	Each panel size corresponds to 60 comoving kpc. 
	White circles indicate the virial radius of the main progenitor. 
	}  
	\label{fig:evolution_color}
\end{figure*}

Fig. \ref{fig:evolution_color} shows the time evolution of the gas density (top), the metallicity (middle) and the temperature (bottom) in the M8run (fiducial run). 
\adc{These snapshots show a typical qualitative picture for the formation sequence of a first galaxy. } 
The \adrr{field of view} is 60 comoving kpc, which roughly corresponds to twice the virial radius of the main progenitor halo at $z = 9$. 
At $z=16$ (left panels), minihaloes form and 
\adc{gas in} some of them collapses via 
molecular cooling, 
resulting in the formation of Pop III stars.
\adc{In the other} minihaloes 
\adc{the gas collapse} is prevented by the LW radiation emitted by nearby Pop III stars formed earlier. 
In \S \ref{sec:LW}, we will discuss 
\adc{in more detail} the impact of LW radiation feedback on the formation of first galaxies. 
At $z=12$ (middle panels), an atomic-cooling halo ($M_{\rm h} \sim 10^{8}~\Msun$) forms. 
At this stage, SNe of Pop II stars generate hot bubbles as seen in the temperature map. 
\adr{The} high thermal pressure of the hot bubble evacuates the surrounding gas and creates a hole-like structure in ISM/IGM. 
\adm{As the galaxy grows, the gravitational potential becomes able to keep the gas against the stellar feedback.
Then, the gas \adc{continues} to accrete onto the galaxy along large scale filaments and is fed to forming stars.}
As the star formation proceeds, the cosmic volume is filled with metals ejected from SNe (middle row). 
In the early stages, Pop III stars pollute the IGM with metals, resulting in a patchy metal-enriched structure. 
At $z = 10$ (right panel), most of the zoom-in region becomes metal-enriched 
\adc{with a volume filling factor of \adm{$\sim 0.4$ ($Z > 10^{-4}~Z_\odot$, within $3r_{\rm vir}$ box region)}.}
 The gas metallicity in the galaxy reaches $Z \gtrsim 10^{-2}~\Zsun$ which is much higher than the critical value for the Pop III-II transition. 

\begin{figure}
	\begin{center}
		\includegraphics[width=8.5 cm,clip]{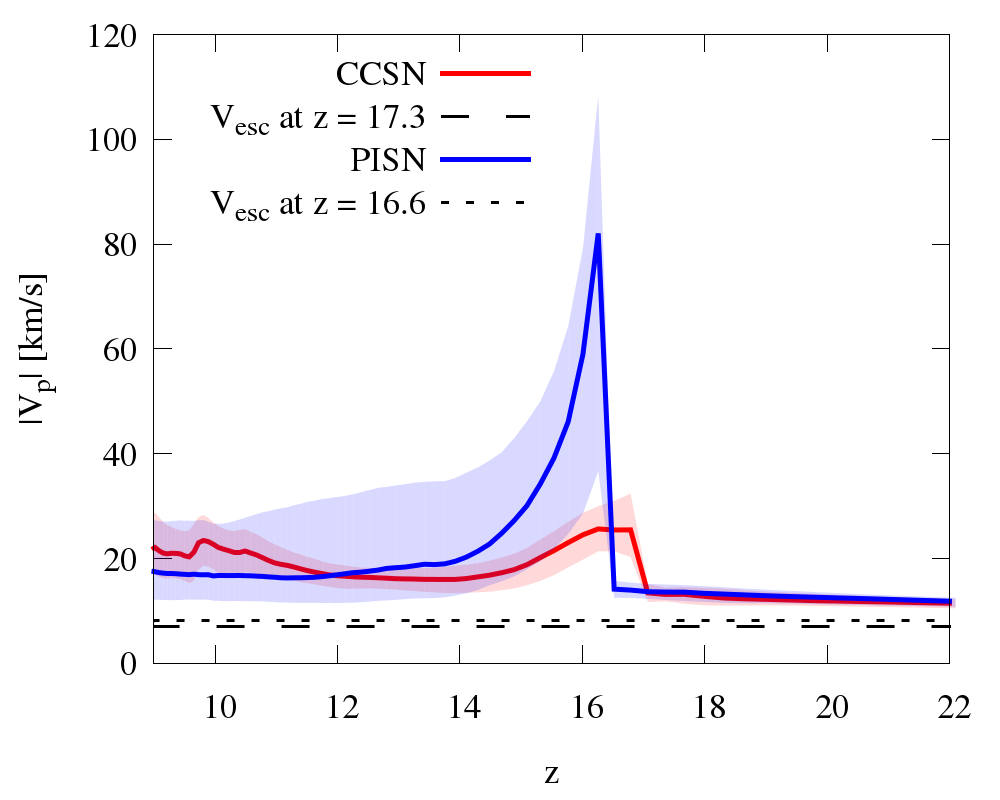}
	\end{center}
	
	\caption{
	Peculiar velocity of gas particles blown out by SN explosions. 
	Solid line is the median and shaded region represents the quartiles. 
	Red line shows the particles which are hosted in a minihalo with Pop III stars that die as CCSN, while blue line is the case for PISN. 
	Dotted/dashed line indicates the escape velocity (Eq. \ref{eq:escape_velocity}) of minihaloes that host the Pop III stars die as CCSN/PISN (\adm{the haloes are selected from fiducial run (M8run) and M8run but assumed top-heavy Pop III IMF, see \S\ref{sec:IMF}}). 
	The redshifts are corresponding to the epoch when the target Pop III stars are born. }
	\label{fig:trajectory}
\end{figure}

To study the propagation of the metal-enriched gas in \adr{the} IGM, we trace the trajectories of all SPH particles in minihaloes hosting Pop III stars. 
Fig. \ref{fig:trajectory} displays the time evolution of the peculiar velocities of the SPH particles in these minihaloes.
\adc{To understand the behaviour seen in Fig. \ref{fig:trajectory}} we select two minihaloes 
\adc{and discuss them in detail}; one minihalo with mass $M_{\rm halo} = 7.1\times 10^{5}~h^{-1}~M_{\odot}$ that hosts a Pop III that will explode in a CCSN at $z \sim 17$ (red line), and another minihalo with $M_{\rm halo} = 1.2\times 10^{6}~h^{-1}~M_{\odot}$ that hosts a  PISN at $z \sim 16 $ (blue line, picked up from M8run \adm{but assumed a top-heavy Pop III IMF, see \S\ref{sec:IMF}}). 
In the former, the CCSN occuring at $z=16.8$ ejects gaseous material with the speed 25.4 km/s. 
On the other hand, the escape velocity of the halo can be evaluated as \citep{Chon+16}: 
\begin{equation}
\label{eq:escape_velocity}
V_{\rm esc} = 7.1~{\rm km~s^{-1}} \left[\frac{\Omega_{\rm m}}{\Omega_{\rm m}(z)}\frac{\Delta}{18\pi^2}\right]^{1/6} \left(\frac{M_{\rm halo}}{10^{6}~h^{-1}~M_{\odot}}\right)^{1/3}\left(\frac{1+z}{10}\right)^{1/2},
\end{equation}
where $\Omega_{\rm m}(z)$ is the 
\adc{matter density} at $z$ and $\Delta$ is the critical overdensity of the halo. 
\adc{This gives the escape velocities around \adm{$\sim 7-8$~km~s$^{-1}$} at $z \sim 17$, smaller than the actual wind velocity. Consequently we find large outflow rates for gas from the haloes.} 
In the case of the Salpeter-like IMF, $\sim 85$ per cent of stars explode as CCSNe. 
This would suggest that, the IGM around most of the minihaloes hosting Pop III stars should be metal-enriched. 
The outflowing gas propagates through the IGM 
\adc{conserving momentum, slowing down as it sweeps up material in the IGM.}
The velocity decreases with 
the distance from its origin as 
\begin{equation}
V \sim 2.7 ~{\rm km~s^{-1}}~ \left(\frac{V_0}{20~\rm{km/s} }\right) \left( \frac{M_{\rm gas}}{10^{5}~\Msun}\right) \left( \frac{D}{1~\rm kpc}\right)^{-3} \left( \frac{1+z}{11}\right)^{-3}, 
\end{equation}
where $V_0$ and $M_{\rm gas}$ are the initial velocity and total mass of the gas blown out from the minihalo, respectively, $D$ is the propagation distance in physical units. 
After traveling sometime in the IGM, the metal-enriched gas eventually falls back onto the halo, which has become more massive after the SN event \citep[see also,][]{Jeon+14}. 
In the case of a PISN, the outflow velocity reaches as high as $\sim 80 ~\rm km ~s^{-1}$. 
As a result, most of the gas reaches far from the host halo and escapes without falling back again. 
We will discuss the impact of the type of Pop III SNe on the properties of the first galaxies later in Sec.\ref{sec:IMF}. 

\begin{figure}
	\begin{center}
		\includegraphics[width=8.5 cm,clip]{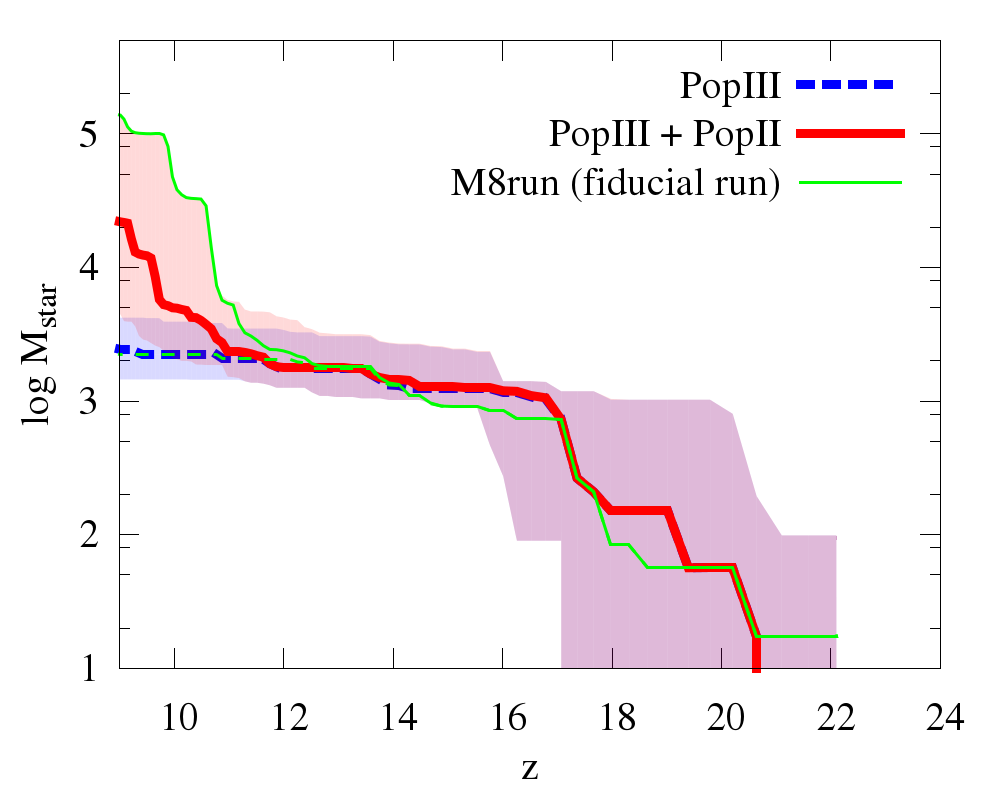}
	\end{center}
	
	\caption{
	Cumulative stellar mass in the five zoom-in regions as a function of redshift. 
	The zoom-in regions are selected so that the halo masses reach $M_{\rm vir} \sim 10^8h^{-1}\solmass$ at $z=9$. 
	In each run, the Salpeter-like IMF is assumed for Pop III stars. 
	\adm{Red solid line shows the total stellar mass formed in Pop III and Pop II, while blue dashed line indicates the total mass formed in Pop III stars.}
	\adm{Thick lines} are the median of five M8runs, and shade depicts the maximum/minimum stellar masses at each redshift bin. 
	\adm{Thin \adr{green} lines correspond to the result of fiducial run (M8run). }
	}
	\label{fig:cumMstar_salpeter}
\end{figure}

Fig. \ref{fig:cumMstar_salpeter} shows the star formation histories in the zoom-in regions. 
A step-like increase of the Pop III cumulative mass corresponds to birth of a single Pop III star. 
We simulate the evolution of five different zoom-in regions that end up with $\sim 10^8~h^{-1}~M_\odot$ haloes at $z=9$ (see table \ref{table:parameter}) to account for statistical variance of such a system. 
At $z \lesssim 12$, Pop III star formation terminates as a result of metal enrichment and Pop II stars start to form continuously in the galaxies thereafter.
Here we define the transition epoch so that the cumulative mass of the Pop II stars exceeds that of Pop III stars. 
\adm{In four out of the five runs the transition epoch is around $z\sim 10$, while in the M8run-3 the mass in Pop II stars falls short of the Pop III stellar mass even at $z = 9$. }
The epochs of the Pop III to II transition are presented in Table \ref{table:parameter}. 

\begin{figure}
	\begin{center}
		\includegraphics[width=9 cm,clip]{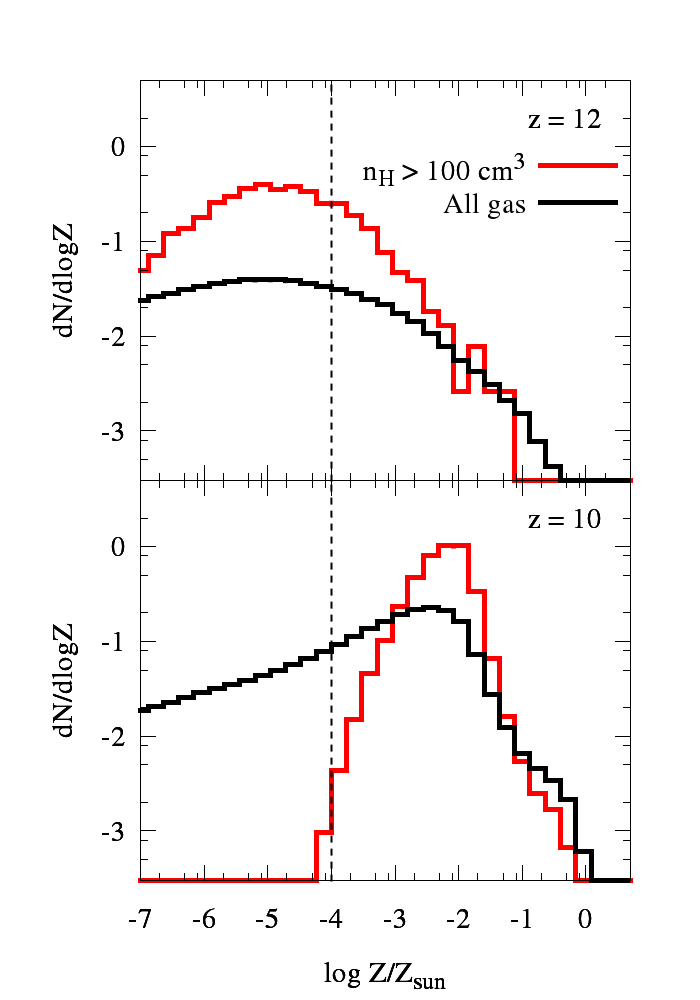}	
	\end{center}
	
	\vspace{-5mm}
	\caption{
	Probability density distribution of the gas metallicity in the zoom-in region at $z=$12 (upper) and 10 (lower).
	The red line depicts the distribution of gas particles that exceed the hydrogen number density of $100~{\rm cm^{-3}}$ while the black line the distribution of all gas particle. 
	\adm{Vertical dashed line indicates the ciritical metallicity $Z_{\rm cr} = 10^{-4}~Z_{\odot}$. }
	}
	\label{fig:metalPDF}
\end{figure}

As seen in Fig. \ref{fig:evolution_color}, most of the zoom-in volume is filled with metal-enriched gas by the time $z \sim 12$.
To understand the transition of the stellar population in more detail, we show the probability density function (PDF) for the gas metallicity in Fig. \ref{fig:metalPDF}. 
At $z=12$ (upper panel), the metallicity has a broad distribution and some gases have such high metallicity as $\sim 0.1~\Zsun$. 
Note also that some dense gases (with $n_{\rm H} > 100~{\rm cm^{-3}}$, solid lines) already have metallicity exceeding the critical value of $Z_{\rm cr} = 10^{-4}~Z_\odot$. 
By $z=10$ (lower panel), most of minihaloes in the zoom-in region have been merged into the main progenitor and most of star formation takes place there. 
At this epoch, more than \textcolor{black}{99} percent of high-density gases have super-critical metallicity. 
Consequently, although gases with sub-critical metallicity still survive in the IGM, \adr{the} mode of star formation completely shifts from Pop III to Pop II. 

\begin{figure}
	\begin{center}
		\includegraphics[width=9.0cm,clip]{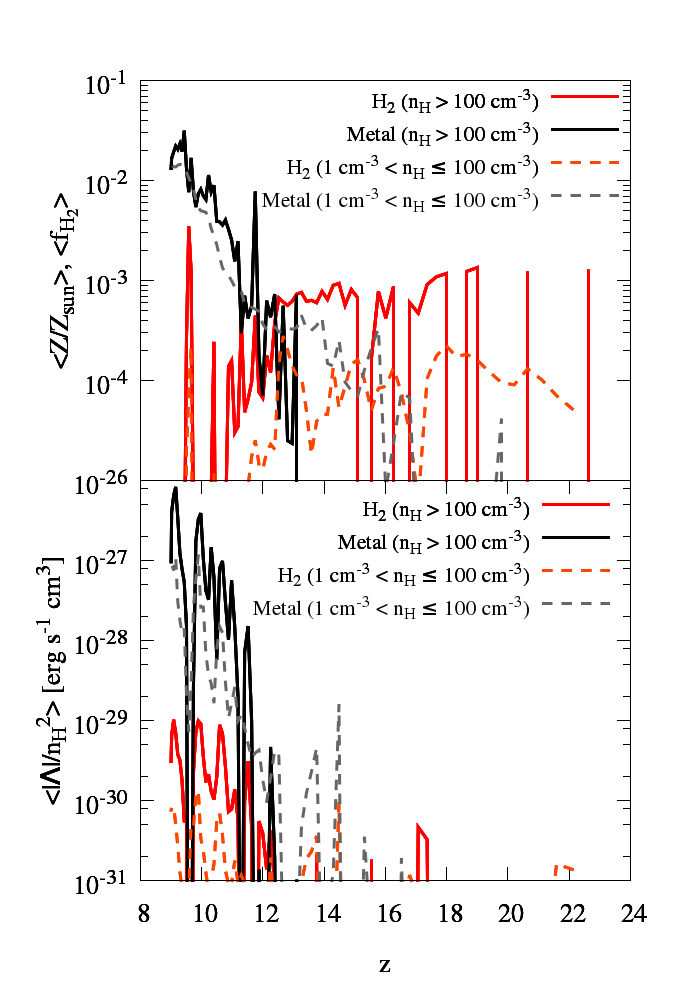}
	\end{center}
	\vspace{-5mm}
	\caption{
	\adm{Time evolution of coolants and the cooling rate in the main-progenitor halo. }
	\adm{The upper panel shows the} evolution of mass-weighted $\HH$ fraction \adm{(red line)} and metallicity \adm{(black line)}. 
	\adm{The lower panel is the mass-weighted cooling rate of molecular hydrogen and the net cooling rate of the metals. }
	\adm{The high-density gas components that exceed the hydrogen number density of $100~{\rm cm^{-3}}$ is depicted by a solid line, while the lower density gas ($1~{\rm cm^{-3}}< n_{\rm H} \leq 100~{\rm cm^{-3}}$) is represented by a dashed line.}
	}  
	\label{fig:H2_metal}
\end{figure}

In \adr{the top panel of} Figure \ref{fig:H2_metal}, the mass-weighted mean $\HH$ fraction and metallicity in high-density gases ($n_{\rm H} > 100~{\rm cm^{-3}}$) in the M8run are shown as a function of redshift.  
\adr{We also display the 
mass-weighted cooling rate of $\HH$ and the net cooling rate of metals in the bottom panel. }
\adr{In early stages before $z \sim 12$, the $\HH$ fraction has a high value of $\langle f_{\rm H_2} \rangle \sim 10^{-3}$ in the high-density regions and  
the $\HH$ is the major coolant.  
Then, at $z \sim 12$, the mean metallicity reaches $10^{-3}~Z_{\odot}$ and the metal cooling becomes as important as that by $\HH$. 
\adr{As the star formation proceeds, the metallicity increases and Pop II stars form mainly via the metal cooling.}  
}

\begin{table*}
\begin{center}
\begin{tabular}{lccccccccc}
\hline \hline
run ID & 
$M_{\rm h}^{z_{\rm end}}$ & $z_{\rm end}$ & $m_{\rm gas}$ & $m_{\rm DM}$ &$M_{\rm h}^{z_{\rm tr}}$ &   $z_{\rm tr}$ & $M_{\rm gas}^{z_{\rm end}}$ & $M_{\rm star}^{z_{\rm end}}$ & $\langle Z \rangle^{z_{\rm end}}$ \\
&$[\Msun/h]$ &  & $[\Msun/h]$  & $[\Msun/h]$  & $[\Msun/h]$ & &$[\Msun/h]$ & $[\Msun/h]$ & $[Z_{\odot}]$   \\
\hline
    M8run (fiducial run) & $1.0\times 10^{8}$ & 9.0 & 12 & 66 & $8.0\times 10^{7}$ & 11.1 & $1.3\times 10^7$ & $7.5\times 10^4$ & $7.4\times 10^{-3}$ \\
    M8run-2  &  $8.8\times 10^{7}$ & 9.0 & 12 & 66 & $8.1\times 10^{7}$ & 9.7 & $6.9\times 10^{6}$ & $1.5\times 10^{4}$ & $3.2\times 10^{-3}$\\
    M8run-3  &   $1.0\times 10^{8}$ & 9.0 & 12 & 66 & - & - & $9.9\times10^{6}$ & $1.6\times10^{3}$ & $8.1\times10^{-4}$ \\
    M8run-4  &  $7.3\times 10^{7}$ & 9.0 & 12 & 66 & $6.7\times 10^{7}$ & 9.5 & $6.9\times10^{6}$ & $1.3\times 10^{4}$ & $1.0\times10^{-3}$\\
    M8run-5  & $1.1\times 10^{8}$  & 9.0 & 12 & 66 &  $7.8\times 10^{7}$ & 10.5 & $1.2\times10^{7}$ & $1.4\times 10^4$ & $1.7\times10^{-3}$\\
    M9run &  $1.2 \times 10^{9}$ &  9.0 & 12 & 66 & $1.4\times 10^{8}$ & 13.3 & $1.5\times10^8$ & $7.7\times10^5$ & $5.7\times10^{-3}$\\
    M9runH & $2.6 \times 10^{9}$& 9.9 & 60 & 330 & $2.9\times 10^{8}$ & 15.5 & $3.5\times10^{8}$ & $3.5\times10^{6}$ & $7.9\times10^{-3}$\\
    M8run-flat & $8.8\times10^{7}$ & 9.0 & 12 & 66 & - & - & $9.5\times 10^{5}$ & 48 & $2.5\times10^{-3}$ \\
    M8run-noLW & $9.5\times10^{7}$ & 9.0 & 12 & 66 & $8.5\times 10^7$ & 10.7 & $6.6\times 10^6$ & $1.4\times10^4$ & $5.2\times10^{-3}$ \\
    M8run-lowZcr  & $1.0\times10^{8}$ & 9.0 & 12 & 66 & $7.5\times10^7$ & 11.3 & $1.1\times10^{7}$ & $6.0\times10^4$ & $4.3\times10^{-3}$
    \\ \hline
\end{tabular}
\end{center}
\caption{$M_{\rm h}^{z_{\rm end}}$: halo mass at the final snapshot in a simulation,  $z_{\rm end}$: redshift at the end of simulation, 
$m_{\rm gas}$: the mass of an SPH particle, $m_{\rm DM}$: the mass of a dark matter particle, $M_{\rm h}^{z_{\rm tr}}$: halo mass at the transition redshift when the dominant population of stars changes from Pop III to Pop II, $z_{\rm tr}$: transition redshift, \adr{$M_{\rm gas}^{z_{\rm end}}$: total gas mass of the halo at the final snapshot, $M_{\rm star}^{z_{\rm end}}$: total stellar mass in the halo at the final snapshot, $\langle Z \rangle^{z_{\rm end}}$: mass-weighted metallicity in the halo at the final snapshot. }}
\label{table:parameter}
\end{table*}

\subsection{Halo mass dependence}
\begin{figure}
	\begin{center}
		\includegraphics[width=8.5 cm,clip]{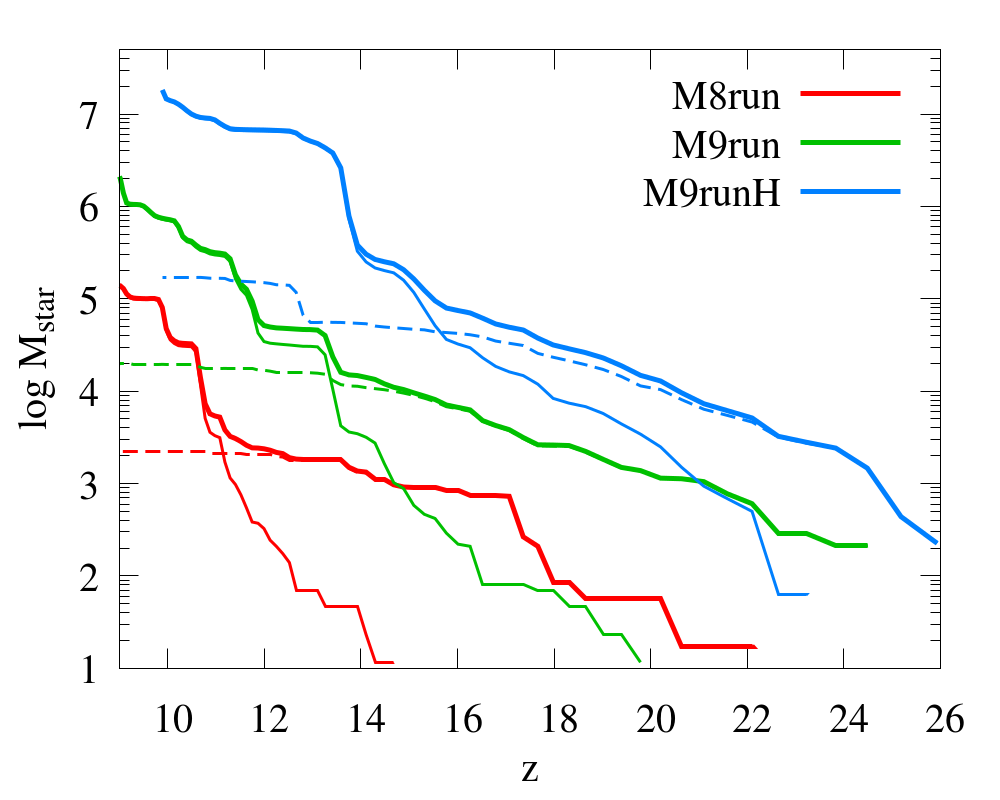}
	\end{center}
	\vspace{-5mm}
	\caption{
	Cumulative stellar mass in the zoom-in region as a function of redshift. 
	Solid \adrr{thick lines show the total stellar masses ever formed.}
	\adrr{Thin dashed and solid lines, respectively, represent the separate contributions by the Pop III and II stars.}
	Colours depict the difference of masses. 
	Red, green and blue respectively corresponds to M8run, M9run and M9runH.   
	}  
	\label{fig:cumMstar}
\end{figure}

\begin{figure}
	\begin{center}
		\includegraphics[width=9.0 cm,clip]{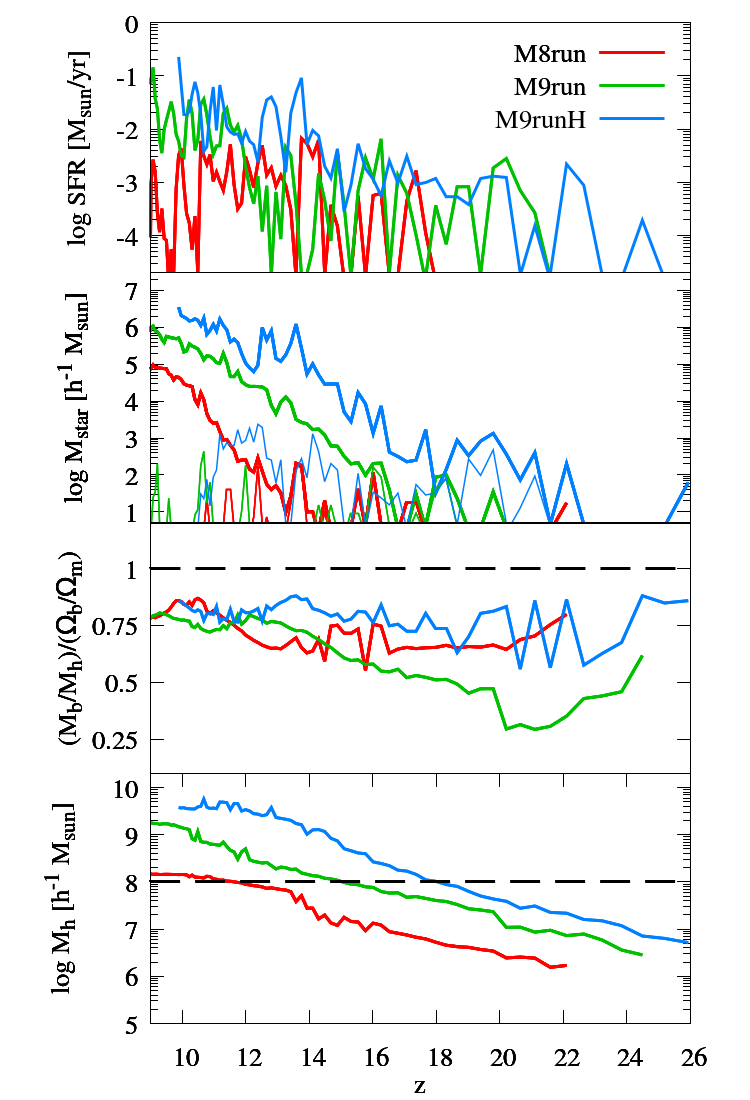}
	\end{center}
	\caption{
	Time evolutions of the main-progenitor halo in M8run (red), M9run (green) and M9runH (blue). 
	\adr{Top panel shows the SFR in the haloes. 
	Second panel indicates the total stellar mass (thick line) and Pop III stellar mass (thin line) in the haloes. }
	Third panel is the baryonic fraction of haloes normalized by the cosmic mean value. 
	Bottom panel displays the growth of the main-progenitor halo masses. 
	\adm{We overplot the lines of barionic fraction equivalent to the cosmic mean in third panel and $M_{\rm h} = 10^{8}~h^{-1}~M_\odot$ in the bottom panel. }
	}  
	\label{fig:prop_progenitor}
\end{figure}

In Fig. \ref{fig:cumMstar}, we compare the star formation histories in three runs, M8run, M9run and M9runH. 
As seen in the figure, the Pop III to II transition redshift tends to be earlier with increasing final halo mass since the halo growth 
\adc{and metal enrichment take place} faster. 
We find that the transition epochs are $z=13.3$ and $15.5$ for M9run and M9runH, respectively (table \ref{table:parameter}). 
The growth of the main progenitors in M8run, M9run and M9runH is shown in Fig. \ref{fig:prop_progenitor}. 
We see in the bottom panel of Fig. \ref{fig:prop_progenitor}, halo masses of the main progenitor in M9run and M9runH reach $\sim10^8~M_\odot$ at $z \sim 15$ and $\sim 18$, respectively. 
\adr{We also see in the second and bottom panels of Fig. \ref{fig:prop_progenitor} that 
the stellar masses rapidly increase once the halo masses reach $\sim 10^{7-8}~M_\odot$ corresponding to the atomic-cooling halo mass.
Because of  deep gravitational potential wells, the atomic-cooling haloes can keep  gas in the face of single SN explosion with $E \sim 10^{51}~\rm erg$.
Therefore, those haloes can form stars continuously with ${\rm SFR} \gtrsim 10^{-3}~\Msunyr$.
One can regard this phase as the birth of the first galaxies. 
In our models, the main-progenitor haloes of M8run, M9run, M9runH continuously show star formation after $z = 14.1,~16.5$ and 21.1, and the corresponding halo masses are $M_{\rm h} = 2.8\times 10^{7}~M_\odot,~5.8\times 10^{7}~M_\odot$ and $3.1\times 10^{7}~M_\odot$, respectively. 
}

\adm{\adc{During the early stages}, the gaseous content (\adr{see third} panel) in the main-progenitor haloes decreases significantly due to star formation and the associated feedback. 
However, the haloes recover the gas as the system grows with time. 
In all the three runs, the baryonic fraction of the first galaxies becomes $\sim 0.75$ of the cosmic mean value by $z$ = 10, and the main progenitor haloes reach the phase in which Pop II stars continue to form \adc{steadily.} }

\begin{figure}
	\begin{center}
		\includegraphics[width=9.0 cm,clip]{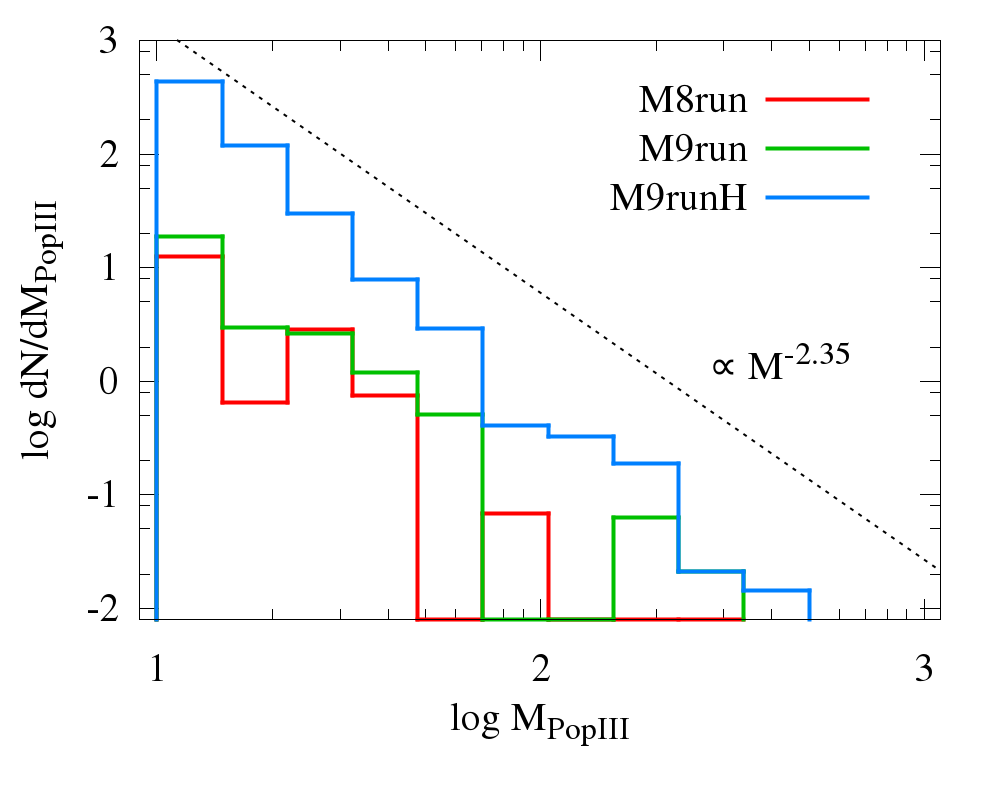}
	\end{center}
	\vspace{-8mm}
	\caption{
	\adr{Mass distribution of all the Pop III stars  formed in the main-progenitor haloes in M8run (red), M9run (green) and M9runH (blue). 
	The dashed line indicates the slope of the assumed IMF ($\propto M^{-2.35}$).}
	}  
	\label{fig:PopIII_mass_spec}
\end{figure}

\adr{Note that, the first galaxies continue to form Pop III stars along with Pop II stars, as shown by the thin line in the second panel of Fig. \ref{fig:prop_progenitor}.
This suggests that metal mixing is not completed and pockets of pristine gas survive. 
In Fig. \ref{fig:PopIII_mass_spec}, we show the mass distribution of all the Pop III stars ever formed in the first galaxies. 
Although the mass spectra almost follow the assumed Pop III IMF, 
no star is formed with a mass higher than $100~M_\odot$ in the M8run
because of the small number of stars.
In the cases of M9run and M9runH, some massive Pop III stars, which eventually explode as either core-collapse or pair-instability SNe, are formed according to the assumed IMF.
Using our Pop III IMF, the released energy per unit stellar mass is estimated as $\sim 8.5\times 10^{49}~{\rm erg~s^{-1}}$, which is $\sim 10$ times larger than that of a Pop II star cluster. 
This suggests the importance of Pop III stellar feedback in controlling the gas dynamics in those haloes if  Pop III stars account for $\gtrsim 10$ percent of the total stellar content. 
Pop III feedback, however, would play a role only in the early stage of galaxy evolution
since the mass fraction in Pop III stars rapidly decreases as metal enrichment proceeds. 
}
\adrr{
Recall that the metal transport among gas particles is not considered in our calculation. More efficient metal mixing would reduce the mass fraction of Pop III stars in the later stage. Our estimate should be regarded as the upper limit in this sense. 
}

As seen in the top-panel in Fig. \ref{fig:prop_progenitor}, star formation occurs intermittently in all three runs.
Such intermittency has been reported for low-mass galaxies by theoretical studies \citep[e.g.,][]{Davis+14,Yajima+17}. 
\adr{This studies claim} that the SN feedback evacuates the gas from the galaxy after starburst, thereby quenching subsequent star formation 
\adc{episode}. 
However, the evacuated gas may fall back to the galaxy after some interval and another episode of starburst may be triggered.
In our simulations, the SFR is higher with increasing halo mass and reaches $\sim 0.1~M_\odot~{\rm yr}^{-1}$ at $z\sim 14$ and $\sim 10$ in M9runH, and $z \sim 9$ in M9run. 
Those starburst galaxies are prime targets for future observations by such telescopes as {\it JWST}. 
We will discuss the observability of the first galaxies later in \S \ref{sec:observability}. 

\section{Effect of different model assumptions}

\subsection{Pop III IMF}\label{sec:IMF}

As seen in \S \ref{sec:formation-evolution}, the gas in the minihalo hosting a Pop III star will be metal-enriched and evacuated by its SN explosion. 
Some of the metal-enriched gas 
\adc{re-accretes} onto the halo.
Therefore, physical conditions in the first galaxies are affected by the statistical nature of Pop III stars \citep{Chen+20}. 
In particular, frequency of SNe strongly depends on the mass spectrum of the Pop III stars. 
However, the IMF of Pop III stars is not firmly constrained yet \citep[e.g.,][]{Susa13,Stacy&Bromm14,Susa+14}.
In this section, we investigate the impact of different Pop III IMFs on the first galaxies. 
In addition to the fiducial Salpeter-like IMF ($\alpha = 2.35$), we also consider a flat IMF with $\alpha=0$ in the same mass range as yet another example  \adm{(M8run-flat)}. 
The event ratios of the CCSNe and PISNe are $N_{\rm {CC}}:N_{\rm {PI}} = 0.85:0.02$ for the Salpeter-like IMF, while $N_{\rm {CC}}:N_{\rm {PI}} = 0.06:0.24$ for the flat IMF. 
The released energy by SNe per unit stellar mass is similar: $8.5\times10^{49}\rm erg/\Msun$ for the Salpeter-like IMF and $9.6 \times 10^{49}~\rm erg/\Msun$ for the flat IMF.

\begin{figure}
	\begin{center}
		\includegraphics[width=8.5 cm,clip]{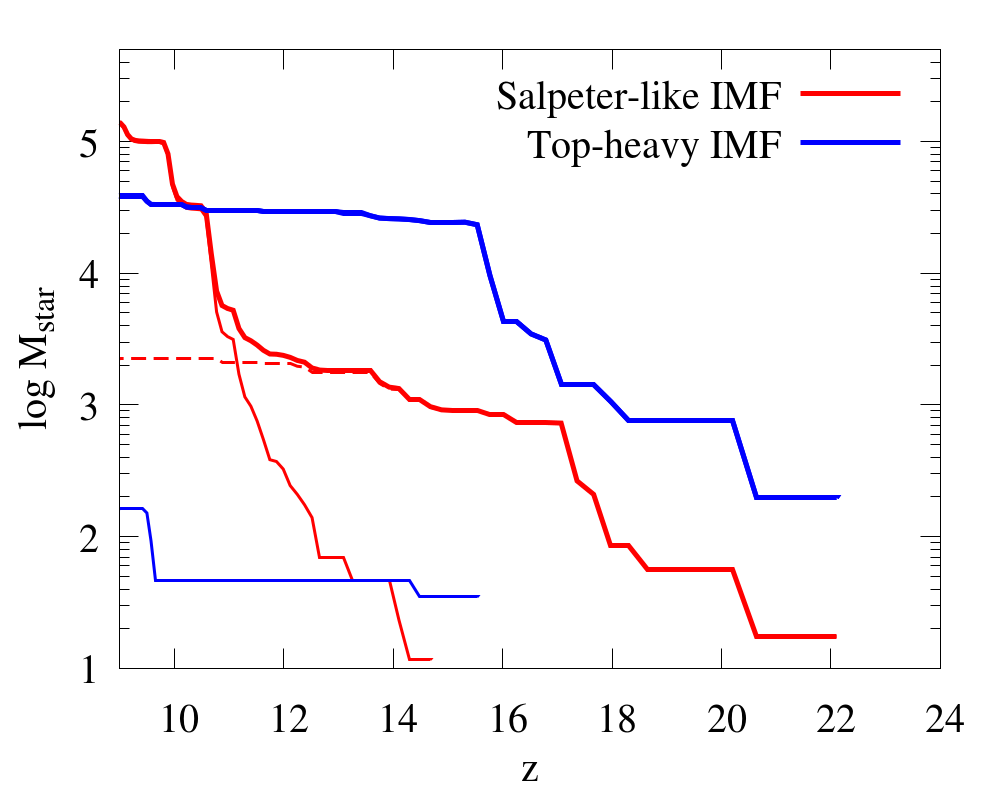}
	\end{center}
	
	\caption{
	Same as Fig. \ref{fig:cumMstar} but for the fidicial run (red) and the top-heavy Pop III IMF run (blue). 
	}
	\label{fig:SFH_IMF}
\end{figure}

Fig. \ref{fig:SFH_IMF} shows the time evolution of the stellar masses for the different IMF models. During the early Pop III phase, the flat IMF run exhibits higher cumulative stellar mass since massive Pop III stars tend to be formed compared to the case with the Salpeter-like IMF. 
However, star formation becomes inefficient (i.e., flattened SFH) at $z \lesssim 16$.
In the flat IMF run, a large fraction of massive Pop III stars explode as PISNe, releasing energy $\approx$ 100 times higher than that of CCSNe. 
In this case, a large amount of gases are evacuated from the host minihaloes with  high velocities of $\gtrsim 50~\rm km~s^{-1}$ as shown in Figure \ref{fig:trajectory}. 
\adr{Part} of the \adr{outflowing} gas sweeps up the surrounding medium and escapes from the gravitational potential well of the haloes. 
Considering that a \adr{fraction} of the SN energy is converted into kinetic energy of the gas in a minihalo, the total momentum is estimated as 
$
	P \sim 1.3 \times 10^7 ~ {\rm \Msun km~s^{-1}}~\left( \frac{\mu}{0.1}\right)^{1/2} \left( \frac{M_{\rm halo}}{10^{6}~\Msun} \right)^{1/2} \left( \frac{E_{\rm SN}}{10^{53}~\rm erg }\right)^{1/2},
$
where $\mu$ is the conversion rate from thermal to kinetic energy.  
The outflowing gas is likely to propagate through \adr{the} IGM conserving the momentum.  
The propagation speed 
\adc{stays at} $v \gtrsim 10~\rm km~s^{-1}$ within $D \sim 1.2~\rm kpc$ at $z \sim 10$ ($\sim 13~\rm kpc$ in comoving scale), which is 
\adc{close to} the circular velocity of the first galaxies at $z \sim 10$. 
In the case \adr{of} the top-heavy IMF model, 31 PISNe explode by $z = 10$. 
Therefore, \adr{the} fallback of the gas within \adr{the} physical volume of $\sim (20~\rm kpc)^{3}$ 
onto the first galaxies is unlikely. 

\begin{figure}
	\begin{center}
		\includegraphics[width=8.5cm,clip]{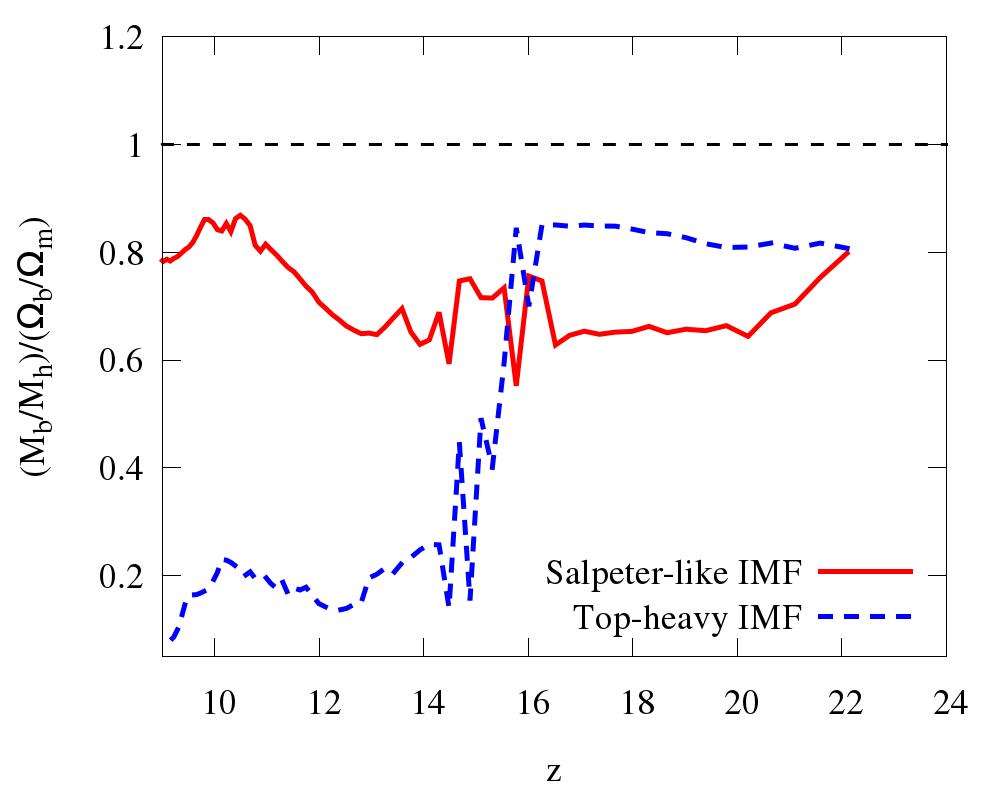}
	\end{center}
	\vspace{-5mm}
	\caption{
	Baryonic fraction in the main-progenitor. 
	\adm{Red solid line depict the result of fiducial run, while blue dashed line indicates the top-heavy Pop III IMF run. }. 
	}  
	\label{fig:baryon_fraction_IMF}
\end{figure}

Fig. \ref{fig:baryon_fraction_IMF} shows the time evolution of the baryonic fraction in the main progenitors relative to the cosmic mean in the cases with the top-heavy and Salpeter IMFs. Because of the evacuation by SNe, the first galaxies
have a low gas content in the case of the top-heavy IMF\adr{, with $\lesssim 0.2$ of the cosmic mean} at $z \sim 10$, 
\adr{while it is \adr{$\gtrsim 0.8$} in the fiducial case.} 
In the top-heavy IMF case, the stellar mass in Pop III stars increases rapidly at $z \sim 15-16$ and reaches more than $10^{4}~\Msun$ (Fig. \ref{fig:SFH_IMF}). Therefore $\gtrsim 20$ PISNe explode and evacuate both the ISM in the halo and surrounding IGM. 
In this simulation, SNe occurs only in less than half the minihaloes either because formed massive Pop III stars directly collapse to BHs without explosion or star formation is prohibited by the LW radiation. 
\adc{Interestingly, we find that multiple PISNe strip gases of nearby minihaloes in which no SN occurs. }
With little Pop II star formation due to a small amount of remaining gas in the galaxy, the Pop III to II transition does not take place by the end of simulation at $z = 9$ in the case of the top-heavy Pop III IMF. 

\begin{figure}
	\begin{center}
		\includegraphics[width=9.2 cm]{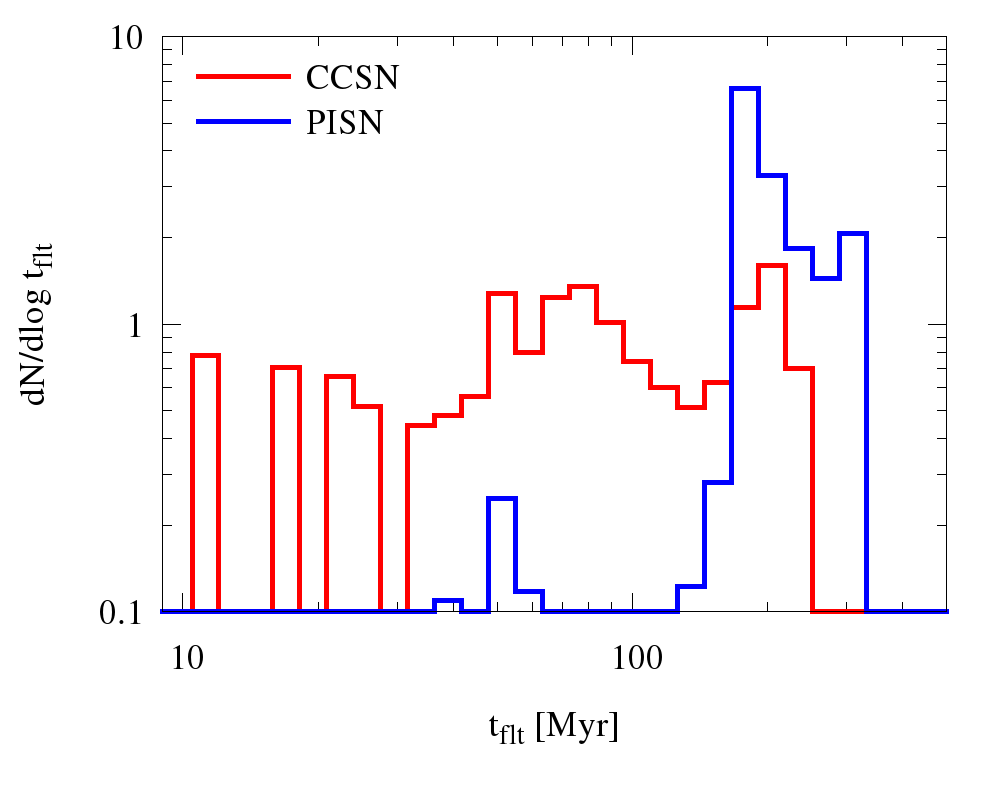}
	\end{center}
	
	\vspace{-5mm}
	\caption{
	Floating time of gas particles ejected from the host halo by SN feedback. 
	The red line indicates the floating time of gas particles ejected via CCSNe in M8run, 
	while the blue line is the gas particles ejected by PISNe in M8run-flat. 
	We pick five Pop III hosting haloes and stack the floating times. 
	}
	\label{fig:floating_time_IMF}
\end{figure}

A small fraction of the gas evacuated from minihaloes 
\adc{re-accretes} onto the galaxy after 
\adc{a long time delay, $t_{\rm flt}$}. 
Figure ~\ref{fig:floating_time_IMF} 
shows the distributions of the floating time of the SPH particles. 
We estimate the floating time by measuring the duration from the time evacuated from minihaloes until they \adc{re-}enter within the virial radius of the (now grown-up) galaxy. \citet{Jeon+14}  showed that the floating time strongly depends on the type of SNe and could be longer than $\sim 100~\rm Myr$ for PISNe. 
Next we see the dependence of the floating time on the shape of \adr{the} IMF.
In the fiducial run, the floating time distributes in a wide range from $\sim 10$ to $\sim 300~\rm Myr$. 
Most of the gas evacuated by CCSNe \adc{re-}accretes onto the galaxy. 
However, there are some variations in the distance from the parental minihaloes to the eventual formation site of the galaxy. This causes the observed wide distribution. 
In addition, a \adr{fraction} of long-traveling gas particles with $\gtrsim 100~\rm Myr$ are due to PISNe. 
With the top-heavy IMF, the distribution concentrates around $\sim 200-400~\rm Myr$. 
This indicates that even the gas in minihaloes locating near the formation site of the galaxy can travel out of its gravitational potential well. 
Later, a \adr{fraction} of the escaping gas \adc{re-}accretes onto the galaxy when the halo mass \adr{reaches} $\sim 10^{8}~\Msun$. 
On the other hand, most of the gas reaches far from the virial radius of the halo at $z \sim 9$, although this material may eventually fall back when the galaxy becomes even more massive at later epoch. 

%
%

\subsection{Lyman-Werner radiation feedback}\label{sec:LW}

\begin{figure}
	\begin{center}
		\includegraphics[width=8.5 cm,clip]{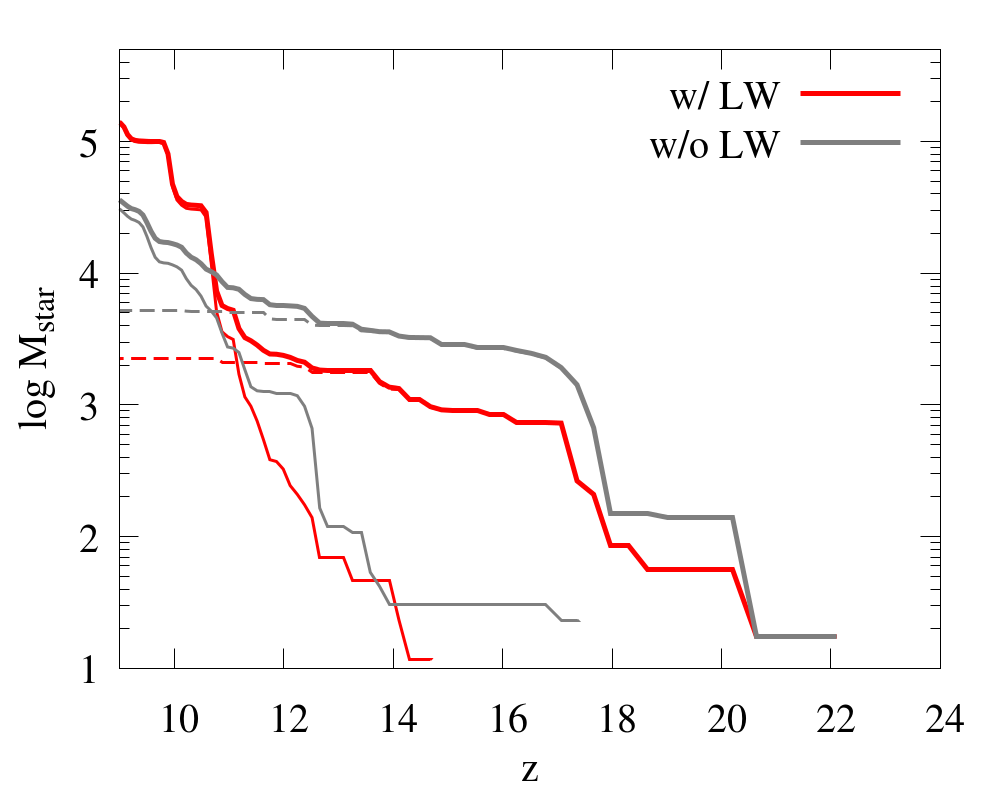}
	\end{center}
	
	\caption{
	Same as Fig. \ref{fig:cumMstar} but for the run with (red line) and without (gray) LW radiation.  
	}
	\label{fig:SFH_LW_comparison}
\end{figure}

\begin{figure}
	\begin{center}
		\includegraphics[width=9cm,clip]{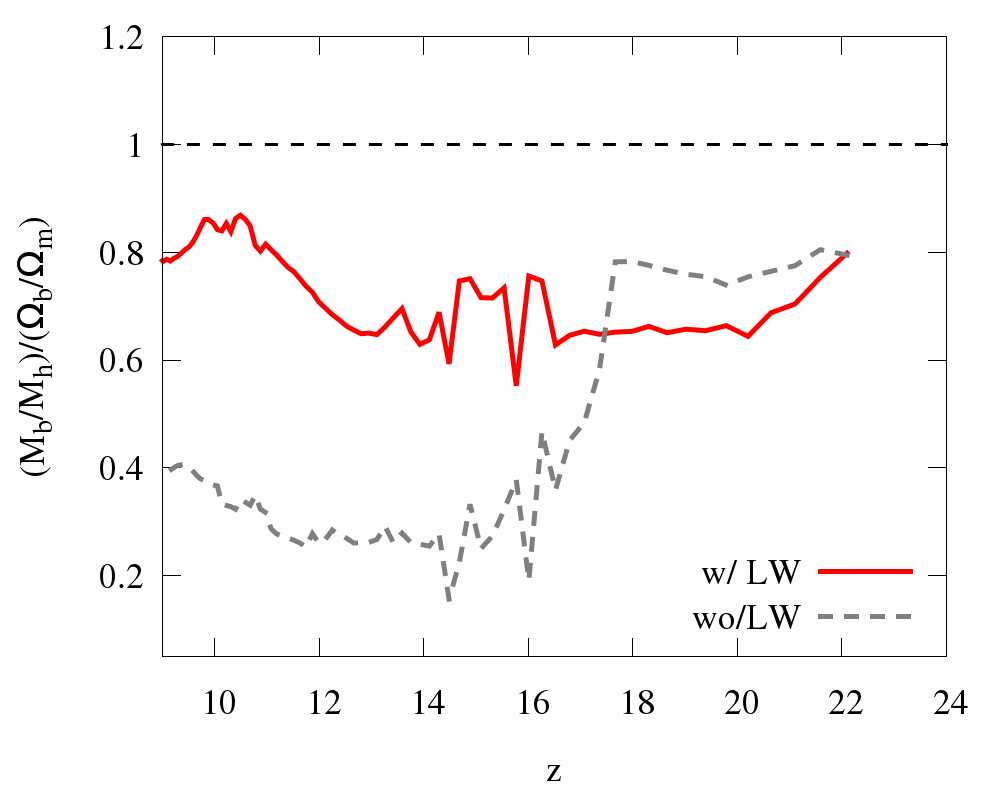}
	\end{center}
	\caption{
	Baryonic fraction relative to the cosmic mean in the main-progenitor. 
	Colours depict the runs with (red) and without (gray) Lyman-Werner radiation feedback. 
	}  
	\label{fig:LW_comparison}
\end{figure}

 LW radiation feedback can play an important role in determining the cosmic star formation history in the early Universe. 
 In particular, the collapse of 
 \adc{gas in} minihaloes is suppressed due to the lack of $\rm H_{2}$ cooling \citep{Haiman+1997}. 
\citet{FiBY} showed that the LW feedback reduces the cosmic SFR density of Pop III stars by a factor of a few. 
Here we investigate the impact of LW feedback on the formation of the first galaxies. 
Figure~\ref{fig:SFH_LW_comparison} shows the time evolution of the total stellar masses. 
We find that the Pop III star formation starts to be suppressed at $z \sim 18$. 
The number of minihaloes hosting Pop III stars is $\sim 50$ percent of that in the case without the LW feedback. 
It is known that star formation is suppressed in minihaloes exposed to the LW flux with $J_{21} \gtrsim 1-10$,  \adrr{where $J_{21}$ is the mean intensity of the background radiation in the unit of $10^{-21}~{\rm erg~cm^{-2}~s^{-1}~Hz^{-1}~sr^{-1}}$ }\citep[e.g.,][]{Latif&Khochfar19, Wise+19}. 
In our model, a Pop III star of $100~\Msun$
\adc{emits} $J_{21} \sim 5.5 ~(D / {\rm 1~kpc})^{-2}$, where $D$ is the physical distance from the star to a neighbouring halo. 
Since the typical separation between minihaloes is $\sim 1~\rm kpc$ at $z \sim 15$, even a single Pop III star can suppress the star formation in neighbouring minihaloes \citep{Agarwal+12, Agarwal+19}. 
Interestingly, once Pop II stars start to form in the first galaxies, the star formation rate becomes higher than that for the the case without the LW feedback, resulting in the total stellar mass twice higher at $z=9$. 
This is induced by more frequent Pop III SNe in the case without the LW feedback 
\citep[see also][]{FiBY}. 

In Fig. \ref{fig:LW_comparison}, we show the time evolution of the baryonic fraction in the main progenitors. 
Without LW feedback, as the Pop III SFR increases at $z \lesssim 18$, the baryonic fraction decreases significantly, resulting in $\lesssim 0.4$ of the cosmic mean. In this case, most of the minihaloes host Pop III stars and $\sim 80$ percent of Pop III stars explode as SNe at the end of their lifetime. 
Therefore most of the minihaloes becomes deprived of the gas while 
\adm{the dark matter continues to accrete onto \adc{them}. }
In the case with LW feedback, the baryonic fraction stays high ($\gtrsim 0.6$) because some gas-rich minihaloes merge with the galaxy. Gas rich 
\adc{infall} induces starbursts at $z \lesssim 12$. 
This shows that the LW feedback is not only important in regulating Pop III star formation, but also has impact on the evolution of the first galaxies.

\subsection{Pop III-II critical metallicity}\label{sec:critical_metallicity}
\begin{figure}
	\begin{center}
		\includegraphics[width=8.5 cm,clip]{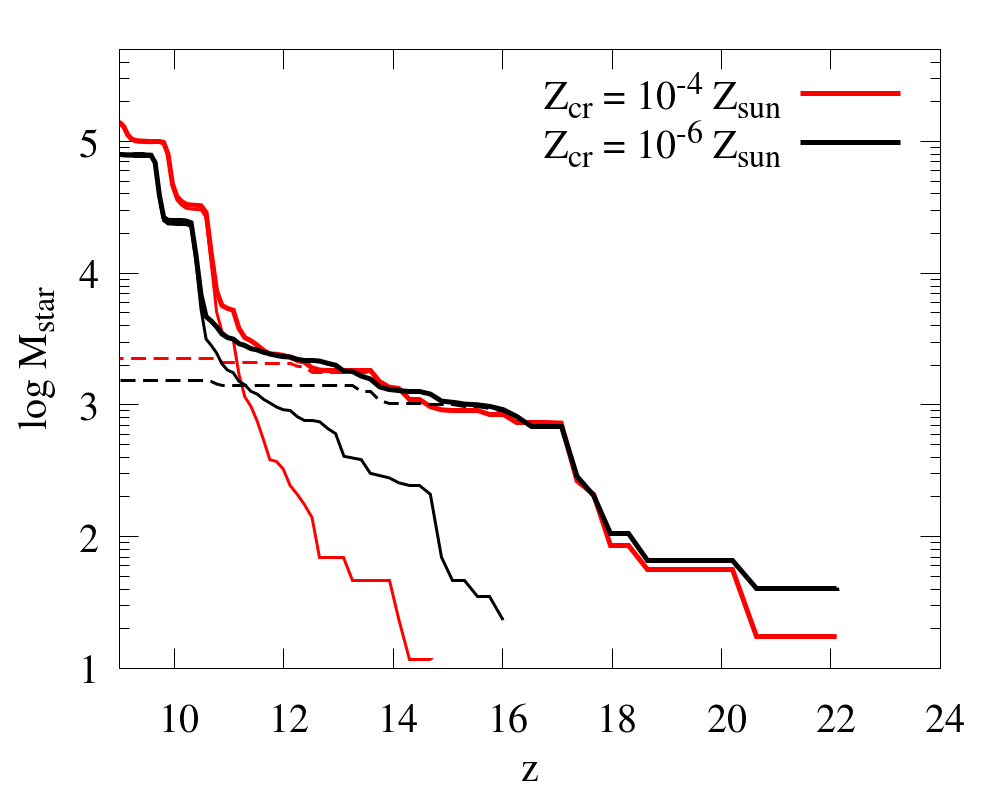}
	\end{center}
	\caption{
	Same as Fig. \ref{fig:cumMstar} but assuming the critical metallicity $Z_{\rm cr} = 10^{-4}~Z_\odot$ (red) and $Z_{\rm cr} = 10^{-6}~Z_\odot$ (black). 
	}  
	\label{fig:cumMstar_Zcrit}
\end{figure}

Thermal evolution of collapsing gas clouds sensitively depends on the metallicity \citep{Omukai2000,Omukai+05, Omukai+2010}. Dust and C/O metals can be efficient coolants and induce fragmentation of the clouds. In a cold metal-enriched gas, the accretion rate onto a formed protostar is lower than in the primordial gas, resulting in the formation of lower-mass stars.
Previous studies have argued that low-mass Pop II-star formation occurs if the metallicity exceeds $Z  \sim 10^{-4}~Z_\odot$ \citep[e.g., ][our fiducial]{Frebel+07, FiBY}, or possibly $Z \sim 10^{-6}~Z_\odot$ due to the efficient dust cooling  \citep{Schneider2003, Schneider2006, Schneider2012}. 

We investigate the impact of the different choices of critical metallicity on the star formation history by performing  additional simulations of the M8run with a critical metallicity of $Z_{\rm cr}=10^{-6}~Z_\odot$ above which Pop II stars form \adr{(M8run-lowZcr)}.
Fig. \ref{fig:cumMstar_Zcrit} shows the star formation histories for the two values of the critical metallicity.
In the case of $Z_{\rm cr}=10^{-6}~\Zsun$, the transition from Pop III to II stars occurs earlier at $z \sim 15$, while the transition redshift is $z \sim 12$ in the fiducial run . 
\adc{This is consistent with the result of numerical simulations by \citet{Maio+10}}. 
However, the difference in the total mass of Pop III stars formed is not significant, and only amounts to $\sim 32$ percent. If a Pop III star ends as a CCSN, the surrounding gas becomes highly metal-enriched. 
Accretion of the metal-enriched gas rapidly enhances the metallicity in the haloes. 
In the fiducial M8run, the time interval of the metallicity increasing from $Z=10^{-6}$ to $10^{-4}~\Zsun$ is quite short $\sim 60~\rm Myr$. 
As shown in the previous section, star formation in the main progenitor proceeds episodically due to stellar feedback. 
\adc{During those $\sim 60~\rm Myr$, only a few star formation events occur.}
This suggests that the star formation history and the formation of the first galaxies are not very sensitive to the detailed value of the critical metallicity \citep{Maio+10}. 

\section{Observability of first galaxies} \label{sec:observability}
First galaxies will be main targets of future observational missions
via UV continuum, \adr{$\rm H\alpha$ and $\rm Ly\alpha$ line emission, especially the Near Infrared Camera (NIRCam), Mid Infrared Instrument (MIRI) and Near Infrared Spectrograph (NIRSpec) on board {\it JWST}.}
Here we estimate the UV continuum and $\rm H\alpha$\adr{/$\rm Ly\alpha$} line fluxes by post-processing and discuss the observability by {\it JWST}.
 
The \adrr{intrinsic} H$\alpha$ \adr{and Ly$\alpha$ luminosities are} calculated by summing up the emissivity of each SPH particle in the galaxy as
\begin{equation}
	L_{\rm H{\alpha}} = \sum_i \frac{m_{\rm i}}{\rho_{\rm i}} j_{\rm H{\alpha}} n_{\rm H_{\rm II}}n_{\rm e}, 
	\label{eq:L_Ha}
\end{equation}
\adr{and
\begin{equation}
	L_{\rm Ly{\alpha}} = \sum_i \frac{m_{\rm i}}{\rho_{\rm i}} j_{\rm Ly{\alpha}} n_{\rm H_{\rm II}}n_{\rm e}, 
	\label{eq:L_Lya}
\end{equation}
}
where $\rho_{i}$ and $m_{\rm i}$ are the gas density and mass of $i$-th SPH particles, $j_{\rm H{\alpha}}$ \adr{and $j_{\rm Ly{\alpha}}$ are} the emission coefficient of \adr{each} line, 
$n_{\rm H_{\rm II}}$ and $n_{\rm e}$ indicate the ionized hydrogen and electron number densities respectively. 
We consider the ionization structure of the ISM calculated on the fly in the simulations.
The spectral energy distribution (SED) of the galaxy is modeled by using {\sc starburst99 }\citep{STARBURST99}. 
First, we generate a table of SEDs for different metallicities ($Z =$0.0004, 0.004, 0.008, 0.02, 0.05) and ages assuming an instantaneous starburst model with the Kroupa IMF \citep{Kroupa2001}. 
We pick the SED of a star particle considering its age and metallicity and compute the integrated SED for the whole galaxy.
\adrr{Here, we do not take into account the intrinsic nebular line emissions.}

Following \citet{Pawlik+11}, 
\adm{we calculate the flux densities of H$\alpha$ \adr{and $\rm Ly\alpha$} lines and UV continuum as} 
\begin{eqnarray}
	f_{\rm H{\alpha}}(\lambda_{\rm obs}) &=& \frac{L_{\rm H{\alpha}}\lambda_{\rm H{\alpha}}(1+z)R}{4\pi cd_{\rm L}(z)^2} \nonumber \\
				&\sim& 60~{\rm nJy}\left(\frac{L_{\rm H\alpha}}{10^{40}~{\rm erg~s^{-1}}}\right)
					\left(\frac{1+z}{10}\right)^{-1}\left(\frac{R}{3000}\right), 
	\label{eq:f_Ha}
\end{eqnarray}
\adr{
\begin{eqnarray}
	f_{\rm Ly{\alpha}}(\lambda_{\rm obs}) &=& \frac{L_{\rm Ly{\alpha}}\lambda_{\rm Ly{\alpha}}(1+z)R}{4\pi cd_{\rm L}(z)^2} \nonumber \\
				&\sim& 3~{\rm nJy}\left(\frac{L_{\rm Ly\alpha}}{10^{40}~{\rm erg~s^{-1}}}\right)
					\left(\frac{1+z}{10}\right)^{-1}\left(\frac{R}{1000}\right), 
	\label{eq:f_Lya}
\end{eqnarray}
}
and 
\begin{eqnarray}
	f_{\rm 1500}(\lambda_{\rm obs})  &=& \frac{L_{\nu}(1+z)}{4\pi cd_{\rm L}(z)^2} \nonumber \\
				&\sim& 1~{\rm nJy}\left(\frac{L_{\nu}}{10^{27}~{\rm erg~s^{-1}~Hz^{-1}}}\right)
					\left(\frac{1+z}{10}\right)^{-1}, 
	\label{eq:f_1500}
\end{eqnarray}
where $c$ is the speed of light, $d_{\rm L}(z)$ is the luminosity distance and $L_{\rm \nu}$ is the luminosity density of UV continuum \adrr{at the wavelength of 1500~\AA}. 

\begin{figure}
	\begin{center}
		\includegraphics[width=9.0 cm,clip]{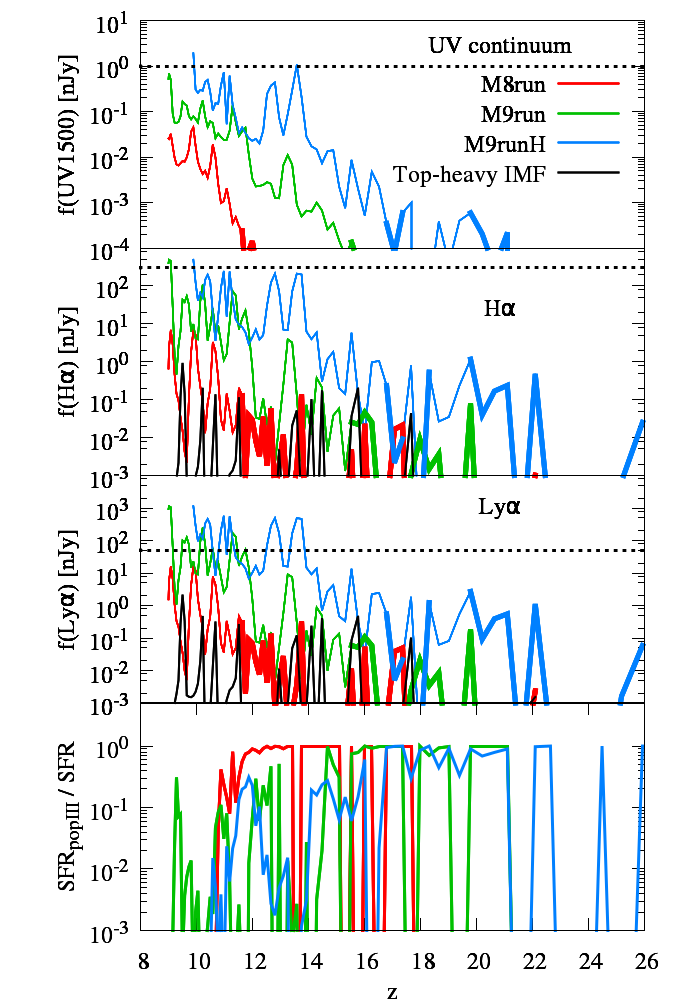}
	\end{center}
	\caption{
	Time evolution of observed flux densities of \adr{the} first galaxies. 
    \adr{Upper, second and third panels show the flux densities of $\rm {H {\alpha}}$ recombination line (Eq. \ref{eq:f_Ha}), UV continuum  (1500\AA, Eq. \ref{eq:f_1500}) and $\rm {Ly {\alpha}}$ recombination line (Eq. \ref{eq:f_Lya}), respectively.}
	Red, \adr{gree and blue} lines are respectively corresponding to M8run, M9run and M9runH. 
	\adr{We also overplot the flux densities in Top-heavy IMF run (\ref{sec:IMF}). }
	Dotted line indicates detection limit of MIRI and NIRCam onboard {\it{JWST}} for exposure time of $10^6$~s and a signal-to-noize ratio S/N=10 \citep{Panagia04,Pawlik+11}. 
	\adr{In the bottom panel, we denote the Pop III star formation rate normalized by the total (Pop III + Pop II) star formation rate. }
	\adrr{In the top three panels, the Pop III dominated phase (SFR of Pop III stars exceeds $>50\%$ of total SFR) is highlighted by the thick lines. }
	}
	\label{fig:flux_density}
\end{figure}

Fig. \ref{fig:flux_density} shows the time evolution of UV continuum, H$\alpha$ and \adr{$\rm Ly\alpha$} flux densities ($f_{\rm 1500}$, $f_{\rm H {\alpha}}$ and \adr{$f_{\rm Ly {\alpha}}$}) of first galaxies with the different masses. 
The \adr{$\rm H\alpha/Ly\alpha$} photons are emitted from H{\sc ii} regions created by young massive stars, while the UV continuum also mainly emitted by the same population.
Therefore, the flux densities in Fig. \ref{fig:flux_density} reflects the instantaneous star formation activity. 
As seen in the Fig. \ref{fig:prop_progenitor}, the star formation in first galaxies takes place intermittently due to the stellar feedback. 
Correspondingly, the flux densities violently fluctuate in time, as also found by \citet{Jeon&Bromm19}. 
The dotted lines in Fig. \ref{fig:flux_density} denote the detection limit for MIRI and NIRCam onboard {{\it JWST}} assuming the exposure time of $10^6$~s and a signal-to-noize ratio S/N = 10. 
\adr{We see in the figure that H$\alpha$ line, Ly$\alpha$ line emissions and UV continuum exceed the detection limit at $z \sim 10$ in the M9runH. }
In this phase, the starburst takes place with SFR $\sim 0.1~M_\odot~{\rm yr^{-1}}$ and the stellar mass is exceeding $\sim 10^{6}~\Msun$. 
To observationally confirm the redshift of the galaxy, 
detection of \adr{H$\alpha$/Ly$\alpha$ lines} 
and UV continuum is preferable.
We suggest that future observations would be able to observe first galaxies with  stellar mass $\gtrsim 10^{6}~\Msun$ beyond redshift $z \sim 10$. 
On the other hand, even JWST cannot detect first galaxies in early phases of their evolution with stellar mass $< 10^{6}~\Msun$. 
\adr{Interestingly, even at $z < 10$, the Pop III star formation rate accounts for $\gtrsim 0.1$ of the total star formation rate in the M9run. 
Thus, Pop III stars make an important contribution to the observable UV fluxes from the first galaxies.}
\adr{As for the Ly$\alpha$ emission, there are uncertainties due to the resonant scattering nature and the IGM transmission \citep{Laursen+11,Laursen+19,ART2,Yajima+14,Yajima+15}. 
In Fig. \ref{fig:flux_density}, we assume that all Ly$\alpha$ photons escape from the galaxies and penetrate into the IGM. 
Before cosmic reionization is completed, development of a giant H{\sc ii} bubble ($r \gtrsim 1~\rm Mpc$) around the galaxy is required for a high IGM transmission \citep{Yajima+18}. 
Thus, shown in Fig. \ref{fig:flux_density} is the upper limit of the emergent Ly$\alpha$ flux. 
Note, however, that we confirm that the flux density exceeds the detection limit at $z \sim 11$ in the M9runH, even if the flux is decreased by multiplying a factor 0.1, corresponding to a low IGM transmission. 
Therefore, the Ly$\alpha$ line might be a powerful tool to detect the first galaxies at $z > 10$.
}
Note that, as described in \S \ref{sec:IMF}, star formation history significantly depends on the Pop III IMF. 
\adr{We overplot the flux densities of the  Top-heavy IMF run in Fig. \ref{fig:flux_density}. 
Due to inactive star formation, the flux densities are significantly lower than that in the fiducial run.
Therefore, future statistical observational studies of the first galaxies might be able to constrain the IMF of Pop III stars. }
In a future work, we plan to investigate the relation between the observational properties of first galaxies and the IMFs of Pop III stars with a larger number of galaxy samples.

\section{\adr{Discussion}}
In this study, we have modelled the formation of individual Pop III stars in minihaloes by taking into account the feedback from Pop III stars in the course of the build-up of first galaxies. 
\adrr{For the photo-ionization feedback, we have estimated the size of an ionized region by the balance between the total recombination rate in the region with the ionizing photon emissivity from the star (Eq \ref{eq:photoheating}).  
This method cannot capture the radiative transfer effect in inhomogeneous density fields, such as the shadowing effect, i.e., shielding of the gas behind a dense clump
from the ionizing radiation \citep[e.g.][]{Susa&Umemura06,Yajima+12DLA}. 
If such an effect is properly included, star formation could continue in the shadowed region.  
}

\adrr{In addition, X-rays from high-mass X-ray binaries (HMXBs) or isolated black-hole (BH) remnants of Pop III stars, which are not taken into account in this work, may change the star formation histories in the first galaxies.}
For instance, \citet{Jeon+12} investigated the impact of  X-ray feedback on the formation of the first galaxies
and found that radiation feedback from isolated BHs and HMXBs suppressed the local star formation significantly. 
Intriguingly, they also observed that the HMXBs enhanced Pop III star formation in nearby minihaloes by inducing  molecular hydrogen formation. 
In \S \ref{sec:LW}, 
we saw that efficient star formation in minihaloes and the resultant frequent SNe from Pop III stars
evacuate the gas and reduce the star formation rate in the first galaxies. 
Some of those Pop III stars would evolve into the HMXBs and X-rays from them can be a key factor to understand the early phase of the first galaxies, although the feedback strength suffers large uncertainties in IMF and binary fraction of Pop III stars.

\adr{
Our current simulations do not have enough resolution to follow the star formation within a small scale of $< 1~\rm pc$, and ab initio calculations of individual star formation as well as binary formation are beyond the scope of this study. 
Instead, we have assumed a simple power-law IMF and determined the mass of individual Pop III stars by random sampling with the weight of the IMF. 
On the other hand, the shape and mass range of the IMF are still under debate and can be different from ours. 
For instance, \citet{Hirano+15} predicted a double-peak IMF with the mass range of $10~M_\odot \lesssim M \lesssim 10^3~M_\odot$.  
With this IMF, PISNe occur more frequently compared to our fiducial simulations and the gas evacuation by the explosions would result in the formation of gas-deprived first galaxies as discussed in \S \ref{sec:IMF}. 
\citet{Hirano+15} also found a relatio between the mass of forming Pop III stars and the local gas infall rate in a minihalo, as well as the external UV radiation field. 
We can evaluate the gas infall rate in a minihalo even with our current resolution and we plan to study Pop III star formation by using more physically motivated relation between the mass and infall rate in a future work. }

\adr{
Recently, \citet{Sugimura+20} investigated the formation of binary/multiple Pop III stars via  fragmentation of a circumstellar disk by performing high-resolution radiation hydrodynamics simulations.
They observed the formation of a Pop III stellar system consisting of a wide binary of 60~$M_\odot$ and 70~$M_\odot$ and smaller mass ($\sim 10~M_\odot$) companion stars . 
If such binary/multiple stellar systems are more common than isolated stars and the IMF is quite different from that assumed here, 
physical properties of the first galaxies can largely differ from our simulations. 
}

\adr{
Currently, even state-of-the-art simulations of galaxy formation cannot resolve individual supernova remnants. 
Therefore, various models of the SN feedback have been proposed so far \citep{Springel05,Stinson+06,DallaVecchia&Schaye12,Kimm&Cen14}.  
Predicted properties of the first galaxies depends on the adopted feedback prescription. 
We have adopted the stochastic thermal feedback model, which successfully produce galactic outflows by overcoming the over-cooling problem \citep[e.g,][]{Katz1996}. 
We observed a clumpy gas structure and intermittent star formation history in the first galaxies \citep[see also][]{Jeon&Bromm19}. 
On the other hand, previous studies with different feedback models instead found a long-standing disk structure and continuous star formation \citep{Pawlik+11, Pawlik+13, Ricotti+16}.
As discussed in Section \ref{sec:observability}, future observations would be able to detect first galaxies only during the starburst phase. 
The duty cycle of the starburst is likely to depend on the feedback process as well as the IMF of Pop III stars. Therefore, future observations combined with theoretical studies might constrain statistically reasonable numerical treatment of SN feedback. 
}

\section{Conclusions}

With cosmological hydrodynamic simulations, we have investigated the formation of the first galaxies that reaches the halo masses of $10^{8-9}M_\odot/h$ at $z=9$ under the influence of feedback from forming Pop III stars. 
The high resolution of the simulations allows us to follow the formation of minihaloes hosting Population III (Pop III) stars and metal enrichment of the inter-galactic medium (IGM) via their supernova (SN) explosions. 
Our major findings can be summarised as follows:
\begin{itemize}
\item The star formation history in the first galaxies depends sensitively on the assumed initial mass function of Pop III stars. 
The dominant stellar population shifts from Pop III to Pop II stars at $z\sim 12-15$ in the case of the power-law Pop III IMF, $dn/dM \propto M^{-2.35}$ with the mass range $10-500~\Msun$. 
\item In the case of a flat top-heavy IMF, frequent pair-instability SNe evacuate the gas from the host halo of a first galaxy. This leads to the formation of a gas-deprived galaxy, where the subsequent Pop II star formation is strongly suppressed.
\item First galaxies are bright in the UV continuum and hydrogen Ly$\alpha$ and H$\alpha$ recombination lines during their starburst phases.   {\it James Webb Space Telescope} will be able to detect both UV, Ly$\alpha$ and H$\alpha$ fluxes from the first galaxies with the halo mass $\gtrsim 10^{9}~\Msun$ at redshift as high as $z \gtrsim 10$. 
\item Lyman-Werner (LW) radiation from Pop III stars affects properties of the first galaxies. 
Without the LW radiation feedback, Pop III stars would be formed in most of the minihaloes, whose gas is evacuated by the subsequent SN explosions, and the total gas mass in the forming galaxy would be $\sim 2-3$ times smaller than that in the case with LW radiation feedback. 
\end{itemize}

In this work, we have shown that physical nature of the first galaxies is affected by previous episodes of Pop III star formation. Upcoming telescopes, including JWST or 30-m class telescopes will be able to detect these first galaxies 
\adc{and shed light on the first generation of stars}. 

%
%
\section*{Acknowledgments}
The authors wish to express their cordial thanks to our most revered mentor, Prof. Masayuki Umemura, President of Astronomical Society of Japan, for his continuous interest and advice. The authors also thank to Ken Ohsuga, Masao Mori and Hajime Fukushima for useful discussions.
The numerical simulations were performed on the computer
cluster, XC50 in NAOJ, and Trinity at Center for Computational Sciences in University of Tsukuba.
This work is financially supported by the Grants-in-Aid for Basic Research by the Ministry of Education, Science and Culture of Japan (HY:17H04827, 20H04724, 21H04489, KO:25287040, 17H01102, 17H02869), 
National Astronomical Observatory of Japan (NAOJ) ALMA Scientific Research Grant Number 2019-11A,  JST FOREST Program, Grant Number JPMJFR202Y (HY).
\adr{CDV acknowledges support through grants RYC-2015-18078 and PGC2018-094975-C22 from the Spanish Ministry of Science and Innovation. }

\section*{Data availability}
The data underlying this article will be shared on reasonable request to the corresponding author. 
%
%
\bibliographystyle{mn}
\bibliography{mn-jour,ref-bibtex}

\label{lastpage}

\end{document}